    \documentclass[fleqn,usenatbib]{mnras}
    
    \usepackage{newtxtext,newtxmath}
    \usepackage{mathrsfs}
    
    \usepackage[T1]{fontenc}
    \usepackage{ae,aecompl}

    \usepackage{graphicx}	
    \usepackage{amsmath}	
    
    \usepackage{amssymb}	
    \usepackage[utf8]{inputenc}
    
    \title[FIR SFRs of six GRB host galaxies with ALMA]{Far-infrared star-formation rates of six GRB host galaxies with ALMA}
    
    \author[Hsiao et al. 2020]{Tiger Yu-Yang Hsiao,$^{1,2}$
    Tetsuya Hashimoto,$^{3,4}$
    Jia-Yuan Chang,$^{5}$
    Tomotsugu Goto,$^{3}$
    \newauthor
    Seong Jin Kim,$^{3}$
    Simon C.-C. Ho,$^{3}$
    Daryl Joe D. Santos,$^{3}$
    Ting-Yi Lu,$^{3}$
    \newauthor
    Alvina Y. L. On,$^{3,4,6}$
    and Ting-Wen Wang$^{3}$
    \\
    $^{1}$Department of Atmospheric Science, National Central University, No.300, Zhongda Rd., Zhongli Dist., Taoyuan City 32001, Taiwan (R.O.C.)\\
    $^{2}$Institute of Astronomy and Astrophysics, Academia Sinica, Taipei 10617, Taiwan (R.O.C.)\\
    $^{3}$Institute of Astronomy, National Tsing Hua University, 101, Section 2. Kuang-Fu Road, Hsinchu, 30013, Taiwan (R.O.C.)\\
    $^{4}$Centre for Informatics and Computation in Astronomy (CICA), National Tsing Hua University,\\
    101, Section 2. Kuang-Fu Road, Hsinchu, 30013, Taiwan (R.O.C.)\\
    $^{5}$Department of Soil and Water Conservation, National Chung Hsing University,\\
    No. 145, Xingda Rd., South Dist., Taichung City 402, Taiwan (R.O.C.)\\
    $^{6}$Mullard Space Science Laboratory, University College London, Holmbury St Mary, Surrey RH5 6NT, UK\\}

    \date{Accepted 2020 June 24. Received 2020 June 24; in original form 2020 March 13}
    
    \pubyear{2020}
    
\begin{document}
    \label{firstpage}
    \pagerange{\pageref{firstpage}--\pageref{lastpage}}
    \maketitle
    
    \begin{abstract}
    Gamma-Ray Bursts (GRBs) can be a promising tracer of cosmic star-formation rate history (CSFRH).
    In order to reveal the CSFRH using GRBs, it is important to understand whether they are biased tracers or not.
    For this purpose, it is crucial to understand properties of GRB host galaxies, in comparison to field galaxies.
    In this work, we report ALMA far-infrared (FIR) observations of six  $z\sim2$ IR-bright GRB host galaxies, which are selected for the brightness in IR.
    Among them, four host galaxies are detected for the first time in the rest-frame FIR.
    In addition to the ALMA data, we collected multi-wavelength data from previous studies for the six GRB host galaxies.
    Spectral energy distribution (SED) fitting analyses were performed with \texttt{CIGALE} to investigate physical properties of the host galaxies, and to test whether active galactic nucleus (AGN) and radio components are required or not.
    Our results indicate that the best-fit templates of five GRB host galaxies do not require an AGN component, suggesting the absence of AGNs.
    One GRB host galaxy, 080207,
    shows a very small AGN contribution.
    While derived stellar masses of the three host galaxies are mostly consistent with those in previous studies, interestingly the value of star-formation rates (SFRs) of all six GRB hosts are inconsistent with previous studies.
    Our results indicate the importance of rest-frame FIR observations to correctly estimate SFRs by covering thermal emission from cold dust heated by star formation.
    
    \end{abstract}
    
    
    \maketitle
    
    
    \section{Introduction}
    \label{introduction}
    Compared to the Short Gamma-Ray Bursts which are believed to be associated with the merger of double compact objects \citep[e.g.,][]{Nakar2007}, 
    Long Gamma-Ray Bursts (hereafter GRBs) are believed to be associated with explosions of massive stars \citep[e.g., ][]{Paczynski1998,MacFadyen1999,Woosley2006}.
    Due to their intense brightnesses and ability to penetrate dusty star-forming regions 
    \citep{Djorgovski2001}, they can be detected at the very distant Universe, i.e., $z \sim 8-9$ \citep[e.g.,][]{Tanvir2009,Salvaterra2009,Cucchiara2011}.
    Therefore, GRBs can be a powerful tracer of cosmic star formation rate history (CSFRH hereafter) 
    \citep[e.g.,][]{Wijers1998,Lamb2000,Porciani2001,Yuksel2008,Trenti2012,Goto2019}.
    For this purpose, it is important to understand what kind of star-forming activities or galaxies are traced by GRBs. 
    
    More than 1,900 GRBs have been detected and well-localised to date
    \footnote[1]{\url{http://www.mpe.mpg.de/~jcg/grbgen.html}} (Note that the total number detected by The Burst And Transient Source Experiment (BATSE) and FERMI is several times larger).
    A number of efforts to observe GRB host galaxies have been made between UV and radio wavelengths over the past decade. For instance, \citet{Hjorth2012} surveyed a large sample of 69 galaxies of the optically unbiased GRB host galaxies (TOUGH) and provided the catalogue. \citet{Japelj2016} studied the influence of the environment on GRB formation and researched SFRs and metallicities of 14 bright GRBs with low-redshift ($z<1$) from {\it SWIFT}/BAT6. 
    \citet{Perley2016} investigated in rest-frame NIR luminosities of 119 GRB host galaxies with a wide range of redshift ($0.03<z<6$). These researches carefully selected the GRB host galaxies in an unbiased way as much as possible.
    Multi-band flux densities of GRB host galaxies selected in this work are reported in previous literature \citep[e.g.,][]{Kuepcue2008,Cucchiara2008,Fugazza2008,Klotz2006,Malesani2006a,Malesani2006b,Ofek2006,Jaunsen2008,Kruhler2011,Hunt2011,Svensson2012,Hjorth2012,Perley2013a,Perley2013b,Hunt2014,Perley2015,Hatsukade2019,Hashimoto2019}.
    Following these observations, physical properties of GRB host galaxies were investigated by fitting the observed spectral energy distributions (SEDs) of GRB host galaxies with galaxy templates \citep[e.g., ][]{Kruhler2011,Perley2013b,Hunt2014,Perley2015,Hashimoto2019}.   

    In general, GRB host galaxies show some distinct characteristics compared to normal star-forming galaxies at least in the low-$z$ Universe. 
    For instance, low-$z$ GRB host galaxies are found to be fainter, bluer, with higher specific star-formation rates and lower metallicities than galaxies that do not host GRBs \citep[e.g., ][]{Christensen2004}.
    Based on the observation of GRB hosts, GRBs averse the environments with high metallicities \citep[e.g.,][]{Stanek2006,Levesque2010,Graham2013,Perley2016}. 
    Although these effects result in the GRBs occuring preferentially in faint galaxies, there are still some GRBs in luminous galaxies.
    \citet{Perley2016} investigated the physical properties such as NIR luminosities of the GRB host galaxies with a wide range of redshifts at $0.03<z<6.3$. Their results indicated that the GRB hosts occurred frequently in faint and low-mass galaxies, while luminous and massive GRB hosts are relatively rare. 
    They also found that dust-obscured GRBs dominate the massive host population but are only rarely seen associated with
    low-mass hosts, reflecting that massive star-forming galaxies are universally and (to some extent) homogeneously
    dusty at high redshift, while low-mass star-forming galaxies retain little dust in their interstellar medium.
    
    The stellar initial mass functions (IMFs) of GRB host galaxies could be different from that in normal star-forming galaxies \citep[e.g., ][]{Hashimoto2018}.
    However, a single IMF shape is conventionally assumed in SED fitting analyses of GRB host galaxies in previous studies \citep[e.g.,][]{Hashimoto2019}.
    A flexibility in the assumed IMFs would be ideal in SED fitting analysis to better understand the physical properties of GRB host galaxies.

    In terms of the requirement to gauge both extincted and unextincted star formation, dust extinction-free methods are especially important because abundant star-forming activity is hidden by dust in general \citep[e.g.,][]{Goto2010}.
    In fact, some GRB host galaxies around $z\sim1-3$ have been localised in ultraluminous infrared galaxies (ULIRGs), indicating intense dust-obscured star formation \citep[e.g.,][]{Perley2017}. 
    These ULIRG host galaxies exhibited highly obscured SFR up to $\sim 10^{3}$ M$_{\odot}$ yr$^{-1}$ \citep[e.g., GRB090404;][]{Perley2017}, 
     probably because
    the fraction of star formation density contributed by ULIRGs increases with redshift \citep[e.g.,][]{Goto2010}.

    After the emission of a gamma-ray burst, there is usually a fading emission in X-ray to radio wavelengths, the so-called \lq afterglow\rq. 
    Radio observation is one of the extinction-free methods that can be utilised to estimate SFRs.
    However, in GRB-host galaxies, long-living radio afterglows might contaminate radio fluxes of the GRB host galaxies 
    for several years after the bursts, which potentially causes overestimation of SFRs when the radio luminosity is used to estimate the star-formation rate
    \citep[e.g.,][]{Murphy2011}. 
    To securely avoid this fundamental problem, radio observations have to be carried out over a longer period of time \citep{Perley2017}.

    Rest-frame FIR observations are also free from dust extinction.
    Afterglows in the rest-frame FIR are negligible after several years from the burst \citep{Hashimoto2019}.
    The rest-frame FIR wavelength is sensitive to the thermal emission from cold dust heated by star-forming regions, which allows us to estimate the SFR more accurately than simply relying on UV to optical or radio data.
    However, most of the previous observations of GRB host galaxies do not include rest-frame FIR data, except for e.g.
    \citet{Berger2003} and \citet{Tanvir2004} which
    presented the FIR observations of host galaxies.
    Even with their efforts \citep[e.g., ][]{Berger2003,Tanvir2004},
    the signal-to-noise ratios in the rest-frame FIR are limited due to the sensitivities of previous telescopes.
    Therefore, SFRs of GRB host galaxies, especially for distant ones, have not been well constrained.
    
    Does the existence of active galactic nucleus (AGN) activity affect the GRB? This has been a matter of debate among previous literature \citep[e.g, ][]{Hatsukade2019, Hashimoto2019}.
    For example, molecular gas excitation of GRB080207 host galaxy is similar to that of excitations caused by AGNs \citep{Hatsukade2019}.
    An optical emission-line diagnostic indicated that the host galaxy is located at a boundary between AGN and \ion{H}{II} regimes \citep{Hatsukade2019}.
    Moreover, the existence of AGN activity correlates with star-formation activity, which could enhance the AGN fraction amongst hosts.
    The presence of AGN affects not only estimates of stellar mass and SFR, but also the IR luminosity. Therefore, it is important to separate the contributions from AGN and SF through the SED fitting.

    In this paper, we present rest-frame FIR detections of six IR-bright GRB host galaxies, as described in Section \ref{sample}. 
    The Atacama Large Millimeter/submillimeter Array (ALMA) bands covered the redshifted FIR continuum of the host galaxies. 
    ALMA's high sensitivity allowed us to detect the rest-frame FIR continuum of these distant GRB host galaxies.
    We measured the flux for each host galaxy at 343.48\,GHz in average. 
    The data including those at rest-frame FIR wavelengths provides a more reliable measurement of obscured SFRs than those without FIR.
    Our sample is divided into two categories, `Group A' that has radio detection and `Group B' that has no radio detection.
    In our SED fitting analyses, we tested with and without the AGN and radio components in configurations for Group A. Due to the lack of radio detection in Group B, we only tested with and without the AGN components in the configurations for Group B (i.e., no radio components for the group B).
    We used an SED fitting code, Code Investigating GALaxy Emission \citep[\texttt{CIGALE}; ][]{Burgarella2005,Noll2009,Boquien2019} which provided two options of IMF:
    \citet{Salpeter1955} and \citet{Chabrier2003}. In this work, we try both IMFs to select the better IMF.
    
    The structure and outline of this paper are as follows.
    We briefly describe our samples in Section \ref{sample}. In Section \ref{data}, we present some configurations and conditions of ALMA archive data and the multi-wavelength photometry. We demonstrate the method and the results of physical parameters
    from SED fitting analyses using \texttt{CIGALE} in Section \ref{analysis_results}. 
    The best fit configurations are discussed and determined using two different IMFs: \citet{Salpeter1955} and \citet{Chabrier2003}, in Section \ref{discussion}. Lastly, we present all four different model settings, the difference, and consistency of physical parameters (SFR, stellar mass, and IR luminosity), in reference to previous works in Section \ref{conclusion}. 
    
    \section{Sample}
    \label{sample}
    In this work, we investigate six IR-bright GRB host galaxies at $z \sim 2$ observed in ALMA band 7, including GRB080207 \citep[$z$ = 2.086;][]{Kruhler2012}, GRB060814 \citep[$z$ = 1.923;][]{Kruhler2012}, GRB070306 \citep[$z$ = 1.496;][]{Jaunsen2008}, GRB081221 \citep[$z$ = 2.260;][]{Salvaterra2012}, GRB071021 \citep[$z$ = 2.452;][]{Kruhler2012} and GRB050915A \citep[$z$ = 2.527;][]{Kruhler2012} host galaxies.
    These host galaxies were selected by ALMA projects 
    2015.1.00927.S and 2016.1.00768.S
    , which aimed at investigating the physical properties of luminous GRB host galaxies  at $z\sim2$. 
    Although
    these galaxies are selected to be IR-bright at high-z
    , this work would be useful for exploring such population of GRB host galaxies. A thorough investigation by \citet{Perley2016} described that dust-obscured GRBs dominate the massive host population but are only rarely seen associated with low-mass hosts. 
    We briefly summarise the previous studies on these individual host galaxies below.
    	
    \subsection{GRB080207 host}

    GRB080207 was first discovered by {\it Swift}/Burst Alert Telescope (BAT) on 2008 February 7 at 21:30:21 UT \citep{Racusin2008}.
    The follow-up observations of deep optical and NIR were performed by \citet{Kuepcue2008}, \citet{Cucchiara2008}, and \citet{Fugazza2008} at about 9 hrs, 9.84 hrs, and 11.3 hrs after the triggers, respectively.
    No afterglow was detected in optical and near-infrared wavelengths \citep[e.g,][]{Hunt2011}.
    The host galaxy of GRB080207 was classified as an extremely red object (ERO) with $R-K=6.3$ and dust-obscured galaxy (DOG) with a flux ratio at $24\,\mu$m to $R-$ band flux $\sim1000$ \citep{Hunt2011}.
 
    The host galaxy has been intensively investigated with optical-to-infrared telescopes such as Keck, Very Large Telescope (VLT), Spitzer and Gemini \citep[e.g.,][]{Hunt2011,Rossi2012,Svensson2012,Hunt2014,Hashimoto2019}.
    The SED fitting analyses of this host galaxy were performed in some previous works based on multi-wavelength photometries.
    \citet{Hunt2014} calculated the following quantities by the SED fitting analysis including the rest-frame FIR continuum detection by {\it Herschel}: star-formation rate of SFR$_{\rm SED}=170.1 ({\rm M}_{\odot} {\rm yr}^{-1})$, total infrared luminosity $\log L_{\rm IR}=12.25 (L_{\odot}),$ total stellar mass $\log M_{*}=11.17 (M_\odot)$, dust mass $\log M_{\rm dust}=8.15(M_\odot)$, dust temperature $T_{\rm dust}=61.3(\rm K)$.
    Additionally, \citet{Hashimoto2019} reported the following results by adding ALMA-detected continuum emission:
    SFR$_{\rm SED}=123.4^{+25.19}_{-21.78}({\rm M}_{\odot} {\rm yr}^{-1})$, $\log L_{\rm IR}=12.26^{+0.05}_{-0.06}(L_{\odot}),$ $\log M_{*}=11.23^{+0.02}_{-0.10}(M_\odot)$  $\log M_{\rm dust}=8.74^{+0.22}_{-0.18}(M_\odot)$, $T_{\rm dust}=39.86^{+1.10}_{-4.11}(\rm K)$.
    The SFRs between these two studies are consistent within uncertainties if $\sim$30\% uncertainty on SFR$_{\rm SED}$ derived by \citet{Hunt2014} is assumed \citep{Hashimoto2019}.
    The stellar masses and the IR luminosities are also consistent within uncertainties.
    
    \subsection{GRB060814 host}
    Prompt emission of GRB060814 was detected by $\it{Swift}$ on 2006 August 14 at 23:02:19 UT \citep{Moretti2006}.
    The afterglow was detected by several telescopes at different times after it was emitted: Télescope à Action Rapide pour les Objets Transitoires (TAROT) after a few minutes \citep{Klotz2006}, VLT in the $R-$band after one hour \citep{Malesani2006a} and $g-$band of Sloan Digital Sky Survey (SDSS) after $\sim 6.7$ hours \citep{Ofek2006}.
    The afterglow was also detected with the United Kingdom Infra-Red Telescope (UKIRT) in the $K-$band approximately 7 hours after the burst \citep{Levan2006}.
    The redshift of GRB060814 host galaxy was measured by \citet{Kruhler2012} to be $z=1.923$ by analysing several nebular emission lines in the near-infrared.
    The host galaxy has been detected in many optical and infrared telescopes such as Keck, VLT, Hubble Space Telescope (HST), Spitzer and Very Large Array (VLA) \citep{Perley2013a,Hjorth2012}.

    Physical properties of this host galaxy were investigated in previous studies \citep{Hjorth2012,Jakobsson2012,Perley2013a,Perley2015}.
    \citet{Perley2013a} and \citet{Perley2015} performed SED fitting analyses of the host galaxy.
    The estimated SFRs are 238 and 209 $M_{\odot}$ yr$^{-1}$, and stellar masses are $9.8\times10^{9}$ and $1.6\times10^{10}$ $M_{\odot}$, respectively.
    
    \subsection{GRB070306 host}
    GRB070306 was first detected at 16:44:28.0 UT on 2007 March 06 \citep{Pandey2007} via {\it Swift}. 
    The afterglow was not detected in optical band with Nordic Optical Telescope (NOT) \citep{Jaunsen2008}.
    \citet{Rol2007} reported that there is a detection of NIR afterglow with William Herschel Telescope (WHT) 3.3 hours later.
    The afterglow was also detected with WHT/Long-slit Intermediate Resolution Infrared Spectrograph (LIRIS) in the $K-$band, suggesting a highly extinguished afterglow \citep{Jaunsen2008}.

    This host galaxy was identified as a heavily obscured star-forming galaxy \citep{Jaunsen2008}, and detected with telescopes such as Spitzer, VLT, {\it Herschel} and VLA \citep{Perley2013a,Hunt2014,Perley2015}.
    The {\it Hershel} data covers the peak wavelength of thermal emission from cold dust in the host galaxy.

    SED fitting analyses of the host galaxy were performed in previous studies.
    \citet{Kruhler2011} reported that $\log(\rm {SFR_{SED}})=1.1^{+0.3}_{-0.2} (M_{\odot}$ yr$^{-1})$  and stellar mass $\log(M_*)=10.39^{+0.19}_{-0.15} (M_\odot)$. 
    With {\it Herschel} data in far-infrared, \citet{Hunt2014} revisited the SED fitting analysis of the host galaxy and derived ${\rm SFR_{SED}}$ $=144.1 (M_{\odot}$ yr$^{-1})$, $\log L_{IR}=12.18 \log L(L_{\odot})$, and $\log M_{*} (M_\odot)=10.05$. 
    \citet{Perley2015} also performed the SED fitting analysis and reported $\rm SFR_{SED}$ $=17^{+7}_{-5} (M_{\odot}$ yr$^{-1})$ and stellar mass $M_{*}=5^{+0.1}_{-0.2}\times10^{10} (M_\odot)$. 
    
    \subsection{GRB081221 host}
    This gamma-ray burst was first detected at 16:21:11 UT on 2008 December 21 with the {\it Swift}/BAT \citep{Hoversten2008}. In addition, the burst event was also detected by Konus-Wind \citep{Golenetskii2008} and the Fermi Gamma-ray Burst Monitor (GBM) \citep{Wilson-Hodge2008}.
    \citet{Malesani2008} and \citet{Afonso2008b} reported that there was no afterglow in optical with 
    NOT and Gamma-Ray Burst Optical/Near-Infrared Detector (GROND) $\sim4$ hrs and 33 hrs after the GRB.
    However the afterglow was detected in NIR by Gemini-N only in the $K-$band 12 hours after the burst was triggered \citep{Tanvir2008}.

    The redshift of the GRB was spectroscopically measured to be $z=2.260$ \citep{Salvaterra2012}. Moreover, the host galaxy was successfully detected in the 
    $B, V, g, I$ and $z$ bands of Keck-I/Low Resolution Imaging Spectrometer (LRIS) \citep{Perley2013a} and $g, r, i$ bands of GROND \citep{Afonso2008,Perley2013a}. The host galaxy was also detected with HST, Gemini-N, Spitzer, and VLA \citep{Perley2013a,Perley2013b,Afonso2008}. 
    
    For further analysis, \citet{Perley2013a} collected the photometric data of the host galaxy and performed SED fitting. They showed  $\rm {SFR_{SED}}$ $=172.8^{+22.8}_{-30.1}(M_{\odot}yr^{-1})$ and $M_{*}=3.7^{+1.1}_{-1.2}\times10^{10}$($M_{\odot}$), respectively. 
    However their photometric data did not include any FIR data.
    
    \subsection{GRB071021 host}
    GRB071021 at 09:41:33 UT on 2007 October 21 was first triggered by {\it Swift} \citep{Sakamoto2007}. 11.25 hours after the burst, the afterglow was detected with Telescopio Nazionale Galileo (TNG) in the $H-$band and $K-$band, but not in the $J-$band \citep{Castro-Tirado2007}.
    
    The redshift of GRB071021 was spectroscopically measured to be $z=2.452$ \citep{Kruhler2012}. 
    The host galaxy of this GRB has been detected in, e.g., $B, V, g, I, z, K$ bands of Keck-I \citep{Perley2013a} and 100, 160, 250, 350, 500 $\mu$m of {\it Herschel} \citep{Hunt2014}.

    \citet{Perley2013a} performed SED fitting to obtain
    $\rm {SFR_{SED}}$ $=190.3^{+25.6}_{-20.3}(M_{\odot}yr^{-1})$ and $M_*= 1.196^{+0.066}_{-0.088}\times10^{11}$($M_{\odot}$).
    
    \subsection{GRB050915A host}
    The prompt emission was detected at 11:22:42 UT on 2005 September 15 by {\it Swift}-BAT \citep{Grupe2005a}. 
    It was followed up by Palomar's 60-inch telescope (P60) and Peters Automated Infrared Imaging Telescope (PAIRITEL). P60 observed the position nine times; the first and last exposure was made 4 minutes and 40 minutes after the trigger, respectively. There was no any detection of the afterglow for all observations of P90 with filters: $R_{\rm C}$, $i$ and $z$ \citep{Cenko2009}. However, the afterglow,  which was actually faint, was detected 11 minutes after the burst by PAIRITEL in the $H-$band \citep{Bloom2005}.
     
    GRB050915A host galaxy was observed and detected by
    e.g., Keck-I, VLT, and Spitzer \citep{Perley2013a,Hjorth2012}. The host galaxy was not detected with the Australia Telescope Compact Array (ATCA) \citep{Michalowski2012}.
    
    SED fitting of GRB050915A host galaxy performed by \citet{Perley2013a} indicated that it is a dust-obscured galaxy with a high star-formation rate of $\rm SFR_{SED}$ $=135.8^{+63.1}_{-48.2}(M_{\odot}yr^{-1})$ and stellar mass of $M_*= 3.67^{+1.6}_{-1.04}\times10^{10}$($M_{\odot}$).

    \section{Data}
    \label{data}
    
    The host galaxies of GRB080207, GRB060814, GRB070306, GRB081221, GRB071021 and GRB050915A were observed with ALMA in projects entitled \lq Luminous, Dust-Enshrouded High-z Galaxies Selected by Gamma-Ray Bursts\rq\ (Project code: 2015.1.00927.S for GRB080207, GRB060814 and GRB071021; and 2016.1.00768.S for GRB070306, GRB081221 and GRB050915A). 
    In each project, three host galaxies were observed. 
    These GRB hosts were observed with ALMA in the band 7 with four spectral windows. The bandwidth is 2\,GHz for each spectral
    window. 
    The host galaxy of GRB080207 was observed with ALMA on June 27, 2016;
    GRB060814 on June 28, 2016;
    GRB070306 on November 26, 2016;
    GRB081221 on December 14, 2016;
    GRB071021 on June 30, 2016;
    and GRB050915A on December 3, 2016. 
    The continuum sensitivities are 0.04 $\mu$Jy and 0.02 $\mu$Jy for projects 2015.1.00927.S and 2016.1.00768.S, respectively. 
    
    The central frequencies of the spectral windows are 336.495, 338.432, 348.495, and 350.495 GHz. The average observed frequency is 343.47925 GHz corresponding to 
    $\lambda_{\rm obs} = 873\,\mu$m.
    The rest-frame wavelengths for GRB080207, GRB060814, GRB070306, GRB081221, GRB071021 and GRB050915A are 282.9, 298.7, 349.8, 267.8, 252.9 and 247.5 $\mu$m, respectively.
    In this work, we use the reduced data provided by the ALMA Science Center.
    
    ALMA arranged these two projects to be observed by a 12-m antenna array.
    The estimated average amounts of precipitable water vapour (PWV) are $\sim$ 1 mm during the observations of GRB080207 and GRB060814 host galaxies, and $\sim$ 0.5 mm for the rest of the host galaxies.
    
\begin{figure*}
    	\includegraphics[width=2.2in]{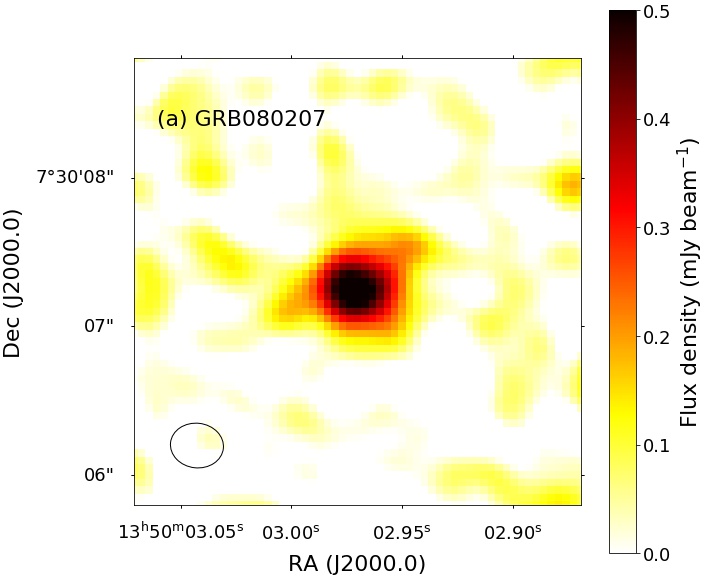}
    	\includegraphics[width=2.2in]{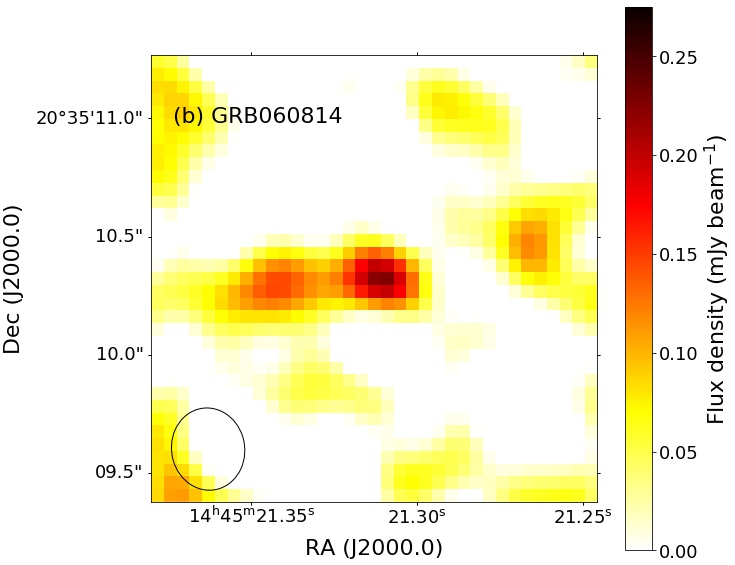}
    	\includegraphics[width=2.2in]{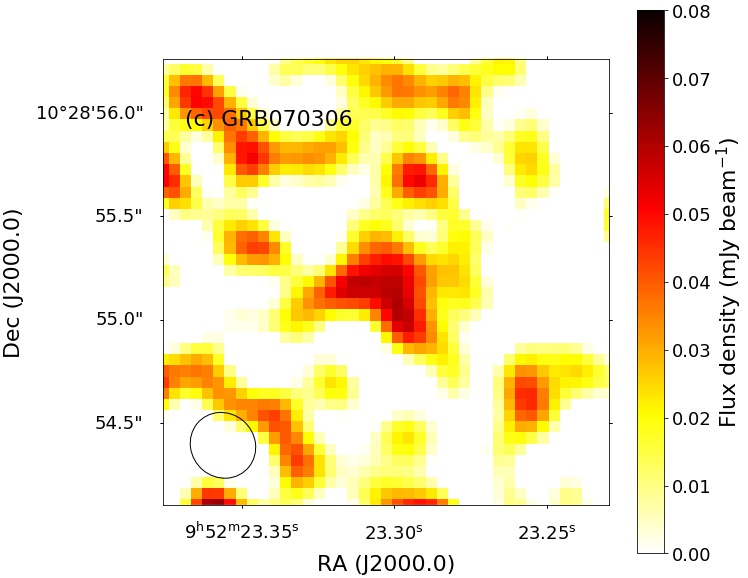}
        \includegraphics[width=2.2in]{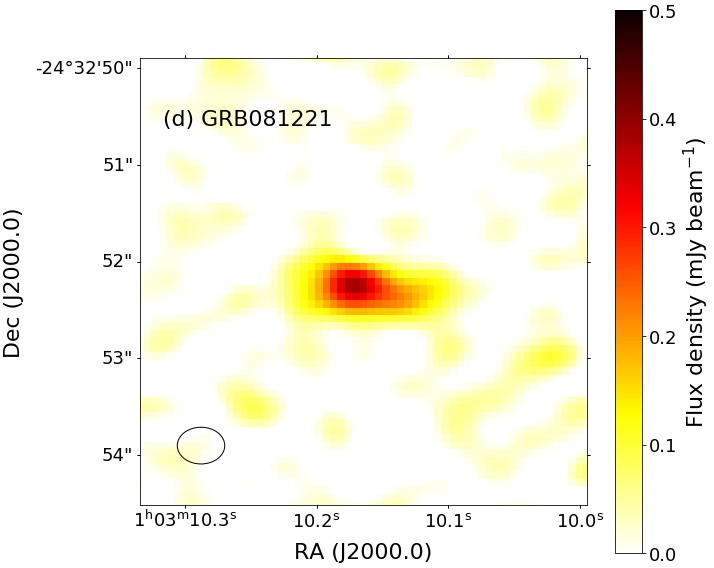}
    	\includegraphics[width=2.2in]{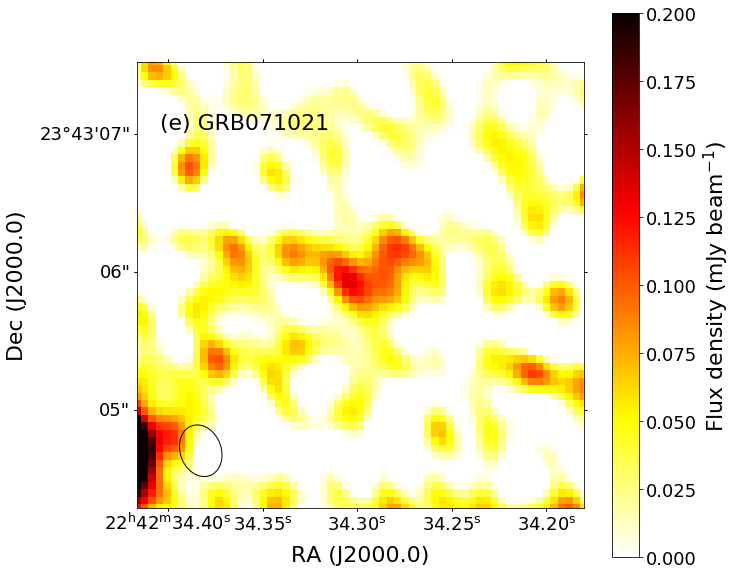}
    	\includegraphics[width=2.2in]{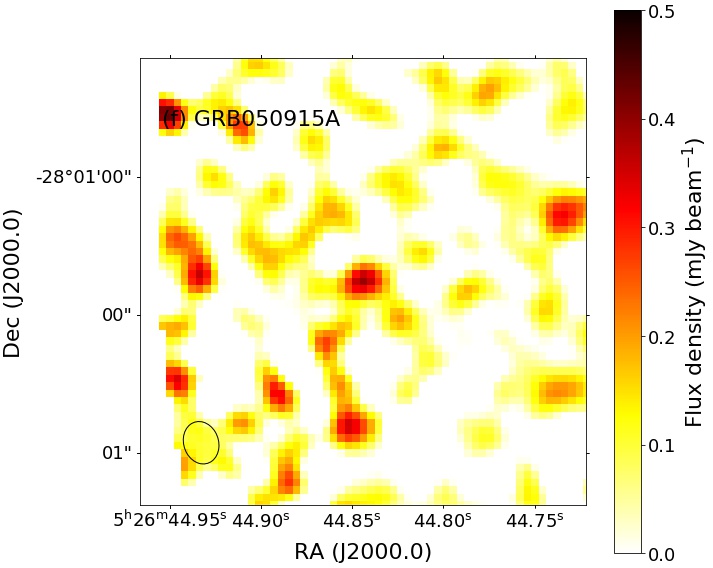}
    	\caption{
       The ALMA band 7 ($873\,\mu$m) images, corresponding to the rest-frame dust continuum and a broad peak in the FIR 
       for the (a) GRB 080207 host, (b) GRB 060814 host, (c) GRB 070306 host, (d) GRB081221 host, (e) GRB071021
    host and (f) GRB050915A host. Beam sizes are indicated by ellipses in the lower left corner of each panel.
        In each panel, the optical/NIR coordinates of the host galaxy are at the center.}
        \label{fig1}
    \end{figure*}

    We detected the rest-frame FIR continuum fluxes of six host galaxies using $\it imfit$ task of CASA version 5.4.0, which allowed us to estimate fluxes, central positions, and their errors. These measurements are shown in Table \ref{tab1}. The rest-frame FIR continuum images are also shown in Fig. \ref{fig1}.
    The detection of GRB080207 (signal-to-noise ratio, S/N = 11), GRB060814 (S/N $\sim 16$) and GRB081221 (S/N $\sim 12$) are at the position of the host galaxies with strong detection. However, the host galaxies of GRB070306, GRB071021 and GRB050915A are marginally detected with S/Ns of $\sim5,\sim4 $ and $\sim3.5$, respectively. 

    We collected UV-to-radio photometric data of six GRB host galaxies reported in previous studies. The multi-wavelength data with the new ALMA data of band 7 presented in this work and references are summarised in Tables \ref{tab2} to \ref{tab7}.

    \begin{table*}
    	\centering
    	\caption{ALMA measurements (band 7) of host galaxies}
    	\label{tab1}
    	\begin{flushleft}
    	\begin{tabular}{|c|c|c|c|}\hline
    	Parameters & GRB 080207 & GRB 060814 & GRB 070306\\ \hline \relax 
    	Flux ($\mu$Jy) & 1430 & 145.3 & 50\\ \relax
    	Uncertainty of flux ($\mu$Jy) &130 &9&11\\ \relax
    	Position of detection (RA) (J2000) &$13:50:02.97 \pm 0.01 $ & $14:45:21.31 \pm 0.01$ &$09:52:23.31 \pm 0.01$\\ \relax
    	Position of detection (Dec) (J2000) & $ +07:30:07.25 \pm 0.01 $& $+20:35:10.32 \pm 0.01$ &$+10:28:55.18 \pm 0.04$\\ \relax
    	Clean beam size in major axis (arcsec) & $0.36$ &$0.35$ &$0.33$\\ \relax
    	Clean beam size in minor axis (arcsec) & $0.30$ &$0.33$ &$0.31$\\ \relax
    	Position angle (degree) & $83.72$ & $17.72$ & $50.50$\\
        \end{tabular}\\

    	\begin{tabular}{|c|c|c|c|}\hline
    	Parameters & GRB 081221 & GRB 071021 & GRB 050915A\\ \hline \relax 
    	Flux ($\mu$Jy) & 735 & 422 & 308\\ \relax
    	Uncertainty of flux ($\mu$Jy) &62 &107 &87\\ \relax
    	Position of detection (RA) (J2000) &$01:03:10.17 \pm 0.01 $ & $22:42:34.3 \pm 0.1 $ &$05:26:44.8 \pm 0.1$\\ \relax
    	Position of detection (Dec) (J2000) & $ -24:32:52.27 \pm 0.01 $& $+23:43:05.9 \pm 0.1$ &$-28:00:59.8 \pm 0.1$\\ \relax
    	Clean beam size in major axis (arcsec) & $0.54$ &$0.39$ &$0.32$\\ \relax
    	Clean beam size in minor axis (arcsec) & $0.38$ &$0.32$ &$0.28$\\ \relax
    	Position angle (degree) & $89.97$ & $28.63$ & $34.39$\\
        \hline
        \end{tabular}\\
       {\bf Notes.} These measurements were calculated by $\it{imfit}$ task in CASA.
        \end{flushleft}
    \end{table*}
    
    \begin{table*}
    	\centering
    	\caption{
    	Multi-wavelength data of GRB 080207 host galaxy.
        }
    	\label{tab2}
    	\begin{flushleft}
    	\begin{tabular}{|l|c|c|c|c|}\hline
    	\multicolumn{5}{|c|}{GRB080207}  \\ \hline \relax
        Observed wavelength ($\mu$m) & Band & Flux density ($\mu$Jy) & Telescope/Instrument & Reference \\ \hline \relax
        0.47068$^{a}$&g&0.04$\pm$0.01&Keck/LRIS & \citet{Svensson2012} \\ 
        0.63755$^{a}$&R&0.093$\pm$0.026&VLT/VIMOS & \citet{Hunt2011} \\ 
        0.75497$^{a}$&I&0.17$\pm$0.05&Keck/LRIS & \citet{Svensson2012} \\ 
        0.95716$^{a}$&z$^{\prime}$&0.35$\pm$0.06&Gemini/GMOS & \citet{Hunt2011} \\ 
        1.12037$^{a}$&F110W&1.75$\pm$0.17&HST/WFC3 & \citet{Svensson2012} \\ 
        1.2&J&1.6$\pm$0.3&VLT/SINFONI & \citet{Hunt2011} \\ 
        1.52791$^{a}$&F160W&2.27$\pm$0.34&HST/NICMOS (NIC3) & \citet{Svensson2012} \\ 
        2.1063$^{a}$&K$^{\prime}$&6.25$\pm$1.62&Gemini/NIRI & \citet{Svensson2012} \\ 
        2.1521$^{a}$&Ks&7.3$\pm$1.0&VLT/ISAAC & \citet{Hunt2011} \\ 
        3.5075$^{a}$&3.6um&14.40$\pm$0.31&Spitzer/IRAC & \citet{Hunt2011} \\ 
        4.4365$^{a}$&4.5um&15.51$\pm$0.44&Spitzer/IRAC & \citet{Hunt2011} \\ 
        5.6281$^{a}$&5.7um&18.53$\pm$1.58&Spitzer/IRAC & \citet{Hunt2011} \\ 
        7.5891$^{a}$&7.9um&12.52$\pm$1.76&Spitzer/IRAC & \citet{Hunt2011} \\ 
        23.21$^{a}$&24um&92.43$\pm$6.50&Spitzer/MIPS & \citet{Hunt2011} \\ 
        97.903$^{a}$&100um&2200$\pm$600&{\it Herschel}/PACS & \citet{Hunt2014} \\ 
        153.94$^{a}$&160um&5900$\pm$1400&{\it Herschel}/PACS & \citet{Hunt2014} \\ 
        242.82$^{a}$&250um&$<$19500$^{c}$&{\it Herschel}/SPIRE & \citet{Hunt2014} \\ 
        340.89$^{a}$&350um&$<$20400$^{c}$&{\it Herschel}/SPIRE & \citet{Hunt2014} \\ 
        450&450um&$<$52483$^{c}$&JCMT/SCUBA2 & \citet{Svensson2012} \\ 
        481.38$^{e}$&band9&11900.0$\pm$1700.0&ALMA & \citet{Hashimoto2019} \\ 
        482.26$^{a}$&500um&$<$21900$^{c}$&{\it Herschel}/SPIRE & \citet{Hunt2014} \\ 
        642.43$^{e}$&band8&3370$\pm$180&ALMA & \citet{Hashimoto2019} \\ 
        850&850um&$<$13183$^{c}$&JCMT/SCUBA2 & \citet{Svensson2012} \\ 
        873$^{b}$&band7&1430$\pm$130&ALMA & This work \\ 
        2103.4&band4&123.0$\pm$24&ALMA & \citet{Hatsukade2019} \\ 
        8137.7&36.83GHz&$<$22.4$^{c}$&VLA & \citet{Svensson2012} \\ 
        57322&5.23GHz&17.1$\pm$2.5$^{d}$&VLA & \citet{Perley2013b} \\ \hline
        \end{tabular}\\
        $^{a}$Effective wavelength retrieved from \url{http://svo2.cab.inta-csic.es/theory/fps3/index.php?mode=browse}.\\
        $^{b}$Central wavelength of the spectral windows used for the continuum image (see Fig. \ref{fig1}a).\\
        $^{c}$3$\sigma$ upper limit, which is not taken into account in the SED fitting analysis.\\ 
        $^{d}$NOT used in the model without radio emission in SED fitting analysis to avoid the possible contaminated flux from the long-lived afterglow. \\
        $^{e}$Central wavelength of the spectral windows calculated by \citet{Hashimoto2019}.
        \end{flushleft}
    \end{table*}
    
    \begin{table*}
    	\centering
    	\caption{
    	Multi-wavelength data of GRB 060814 host galaxy.
        }
    	\label{tab3}
    	\begin{flushleft}
    	\begin{tabular}{|l|c|c|c|c|}\hline
    	\multicolumn{5}{|c|}{GRB060814}  \\ \hline \relax
        Observed wavelength ($\mu$m) & Band & Flux density ($\mu$Jy) & Telescope/Instrument & Reference \\ \hline \relax
    
        0.54037$^{a}$&V&2.27$\pm$0.22&Keck/LRIS & \citet{Perley2013a} \\ 
        0.6550&R&2.46$\pm$0.26&VLT/FORS2 & \citet{Perley2013a},\citet{Hjorth2012} \\ 
        0.75497$^{a}$&I&2.91$\pm$0.37&Keck/LRIS & \citet{Perley2013a} \\ 
        1.23646$^{a}$&F125W&5.32$\pm$0.79&HST/WFC3 & \citet{Perley2013a} \\ 
        1.52791$^{a}$&F160W&7.28$\pm$1.08&HST/WFC3 & \citet{Perley2013a} \\ 
        2.1600&K&6.5$\pm$0.6&VLT/ISAAC & \citet{Perley2013a},\citet{Hjorth2012}\\ 
        3.5075$^{a}$&3.6&10.19$\pm$0.88&Spitzer/IRAC & \citet{Perley2013a} \\ 
        4.4365$^{a}$&4.5&12.25$\pm$0.82&Spitzer/IRAC & \citet{Perley2013a} \\ 
        873$^{b}$&band7&145.3$\pm$9.0&ALMA & This work \\ 
        100000&S/3GHz&11.34$\pm$3.12$^{d}$&VLA & \citet{Perley2015} \\ 
        210000& 21cm &$<430^{c}$& WSRT & \citet{Michalowski2012} \\ \hline
        
        \end{tabular}\\
        $^{a}$Effective wavelength retrieved from \url{http://svo2.cab.inta-csic.es/theory/fps3/index.php?mode=browse}.\\
        $^{b}$Central wavelength of the spectral windows used for the continuum image (see Fig. \ref{fig1}b).\\
        $^{c}$3$\sigma$ upper limit, which is not taken into account in the SED fitting analysis.\\ 
        $^{d}$NOT used in the model without radio emission in SED fitting analysis to avoid the possible contaminated flux from the long-lived afterglow.  
        \end{flushleft}
    \end{table*}
    
    \begin{table*}
    	\centering
    	\caption{
    	Multi-wavelength data of GRB 070306 host galaxy.
        }
    	\label{tab4}
    	\begin{flushleft}
    	\begin{tabular}{|l|c|c|c|c|}\hline
    	\multicolumn{5}{|c|}{GRB070306}  \\ \hline \relax
        Observed wavelength ($\mu$m) & Band & Flux density ($\mu$Jy) & Telescope/Instrument & Reference \\ \hline \relax
        0.35949$^{a}$&u&2.24$\pm$1.18&SDSS & \citet{Perley2013a}, \citet{Jaunsen2008} \\ 
        0.45045$^{a}$&g&2.78$\pm$0.24&GROND & \citet{Perley2013a}, \citet{Kruhler2011} \\
        0.46404$^{a}$&g&2.4$\pm$0.41&SDSS & \citet{Perley2013a}, \citet{Jaunsen2008} \\ 
        0.60980$^{a}$&r&2.29$\pm$0.2&GROND & \citet{Perley2013a}, \citet{Kruhler2011} \\ 
        0.61223$^{a}$&r&2.56$\pm$0.66&SDSS & \citet{Perley2013a}, \citet{Jaunsen2008} \\ 
        0.6550&R&2.46$\pm$0.21&VLT/FORS2 & \citet{Perley2013a}, \citet{Kruhler2011} \\
        0.76047$^{a}$&i&2.88$\pm$0.37&GROND & \citet{Perley2013a}, \citet{Kruhler2011}\\ 
        0.84669$^{a}$&I&3.42$\pm$0.65&NOT/ALFOSC & \citet{Perley2013a}, \citet{Jaunsen2008} \\ 
        0.89293$^{a}$&z&2.71$\pm$0.46&GROND & \citet{Perley2013a}, \citet{Kruhler2011} \\
        1.23646$^{a}$&F125W&6.4$\pm$0.18&HST/WFC3 & \citet{Perley2013a} \\ 
        1.2500&J&8.37$\pm$0.64&VLT/ISAAC & \citet{Perley2013a}, \citet{Kruhler2011} \\ 
        1.52791$^{a}$&F160W&7.79$\pm$0.22&HST/WFC3 & \citet{Perley2013a} \\ 
        1.63302$^{a}$&H&9.26$\pm$0.35&GROND & \citet{Perley2013a}, \citet{Kruhler2011} \\
        1.65&H&12.21$\pm$1.43&VLT/ISAAC & \citet{Perley2013a}, \citet{Kruhler2011}\\ 
        2.16&K&10.29$\pm$0.99&VLT/ISAAC & \citet{Perley2013a}, \citet{Kruhler2011}\\ 
        3.5075$^{a}$&3.6&10.65$\pm$0.48&Spitzer/IRAC & \citet{Perley2013a} \\ 
        4.4365$^{a}$&4.5&12.28$\pm$0.59&Spitzer/IRAC & \citet{Perley2013a} \\ 
        97.903$^{a}$&100$\mu m$&4900$\pm$700&{\it Herschel}/PACS & \citet{Hunt2014} \\ 
        153.94$^{a}$&160$\mu m$&10700$\pm$2000&{\it Herschel}/PACS & \citet{Hunt2014} \\ 
        873$^{b}$&band7&50$\pm$11&ALMA & This work \\ 
        100000&S/3GHz&11.34$\pm$2.84$^{c}$&VLA & \citet{Perley2015} \\ \hline
        
        \end{tabular}\\
        $^{a}$Effective wavelength retrieved from \url{http://svo2.cab.inta-csic.es/theory/fps3/index.php?mode=browse}.\\
        $^{b}$Central wavelength of the spectral windows used for the continuum image (see Fig. \ref{fig1}c).\\
        $^{c}$NOT used in the model without radio emission in SED fitting analysis to avoid the possible contaminated flux from the long-lived afterglow.  
        \end{flushleft}
    \end{table*}
    \begin{table*}
    	\centering
    	\caption{
    	Multi-wavelength data of GRB 081221 host galaxy.
        }
    	\label{tab5}
    	\begin{flushleft}
    	\begin{tabular}{|l|c|c|c|c|}\hline
    	\multicolumn{5}{|c|}{GRB081221}  \\ \hline \relax
        Observed wavelength ($\mu$m) & Band & Flux density ($\mu$Jy) & Telescope/Instrument & Reference \\ \hline \relax
    
        0.43221$^{a}$& B & 0.24$\pm$0.03& Keck-I/LRIS & \citet{Perley2013a} \\
        0.45045$^{a}$& g & 0.32$\pm$0.1& GROND & \citet{Afonso2008}\\ 
        0.47068$^{a}$&g& 0.36$\pm$0.06& Keck-I/LRIS & \citet{Perley2013a}\\ 
        0.54037$^{a}$&V& 0.33$\pm$0.05& Keck-I/LRIS & \citet{Perley2013a}\\ 
        0.60980&r&0.5$\pm$0.1& GROND & \citet{Perley2013a} \\
        0.75497$^{a}$&I& 0.64$\pm$0.2&Keck-I/LRIS & \citet{Perley2013a} \\ 
        0.76047$^{a}$&i& 0.65$\pm$0.29&GROND & \citet{Perley2013a} \\ 
        1.03264$^{a}$&z& 0.96$\pm$0.14&Keck-I/LRIS & \citet{Perley2013a} \\ 
        1.04317$^{a}$&F105W& 1.08$\pm$0.07&HST/WFC3 & \citet{Perley2013a} \\ 
        1.52791$^{a}$&F160W& 2.19$\pm$0.08&HST/WFC3 & \citet{Perley2013a} \\ 
        2.2003$^{a}$&K& 3.66$\pm$0.46&Gemini-N/NIRI & \citet{Perley2013a} \\ 
        3.5075$^{a}$&3.6& 8.02$\pm$0.77&Spitzer/IRAC & \citet{Perley2013a} \\ 
        4.4365$^{a}$&4.5& 9.2$\pm$0.89&Spitzer/IRAC & \citet{Perley2013a} \\ 
        873$^{b}$&band7&735$\pm$62&ALMA & This work \\ 
        57322&5.23GHz&<17.2$^{c}$&VLA&\citet{Perley2013b} \\
        \hline 
        \end{tabular}\\
        $^{a}$Effective wavelength retrieved from \url{http://svo2.cab.inta-csic.es/theory/fps3/index.php?mode=browse}.\\
        $^{b}$Central wavelength of the spectral windows used for the continuum image (see Fig. \ref{fig1}d).\\
        $^{c}$3$\sigma$ upper limit, which is not taken into account in the SED fitting analysis.
        \end{flushleft}
    \end{table*}
    \begin{table*}
    	\centering
    	\caption{
    	Multi-wavelength data of GRB 071021 host galaxy.
        }
    	\label{tab6}
    	\begin{flushleft}
    	\begin{tabular}{|l|c|c|c|c|}\hline
    	\multicolumn{5}{|c|}{GRB071021}  \\ \hline \relax
        Observed wavelength ($\mu$m) & Band & Flux density ($\mu$Jy) & Telescope/Instrument & Reference \\ \hline \relax
    
        0.43221$^{a}$& B & 0.13$\pm$0.03& Keck-I/LRIS & \citet{Perley2013a} \\
        0.47068$^{a}$&g& 0.21$\pm$0.02& Keck-I/LRIS & \citet{Perley2013a}\\ 
        0.54037$^{a}$&V& 0.33$\pm$0.06& Keck-I/LRIS & \citet{Perley2013a}\\ 
        0.75497$^{a}$&I& 0.71$\pm$0.14&Keck-I/LRIS & \citet{Perley2013a} \\ 
        1.03264$^{a}$&z& 0.49$\pm$0.16&Keck-I/LRIS & \citet{Perley2013a} \\  
        1.04317$^{a}$&F105W& 0.75$\pm$0.07&HST/WFC3 & \citet{Perley2013a} \\ 
        1.24581$^{a}$&J& $<1.35^{c}$&Gemini-N/NIRI & \citet{Perley2013a} \\
        1.52791$^{a}$&F160W& 1.93$\pm$0.07&HST/WFC3 & \citet{Perley2013a} \\ 
        2.18186$^{a}$&K& 5.42$\pm$1.28&Keck-I/NIRC2 & \citet{Perley2013a} \\ 
        3.5075$^{a}$&3.6& 10.33$\pm$0.4&Spitzer/IRAC & \citet{Perley2013a} \\ 
        4.4365$^{a}$&4.5& 12.26$\pm$0.47&Spitzer/IRAC & \citet{Perley2013a} \\ 
        873$^{b}$&band7&422$\pm$107&ALMA & This work \\ 
        97.903$^{a}$&$100\mu m$&<6000$^{c,d}$&{\it Herschel}/PACS & \citet{Hunt2014} \\ 
        153.94$^{a}$&$160\mu m$&<19700$^{c,d}$&{\it Herschel}/PACS & \citet{Hunt2014} \\ 
        242.84$^{a}$&$250\mu m$&$<$19800$^{c,d}$&{\it Herschel}/SPIRE & \citet{Hunt2014} \\ 
        340.89$^{a}$&$350\mu m$&$<$16200$^{c,d}$&{\it Herschel}/SPIRE & \citet{Hunt2014} \\ 
        482.26$^{a}$&$500\mu m$&$<$21900$^{c,d}$&{\it Herschel}/SPIRE & \citet{Hunt2014} \\ 
        2861&104.8GHz&<510$^{c}$&Plateau de Bure Interferometer (PdBI)&\citet{Postigo2012} \\
        3486&86GHz&<450$^{c}$&PdBI&\citet{Postigo2012} \\
        57322&5.23GHz&<25.4$^{c}$&VLA&\citet{Perley2013b} \\
        \hline 
        \end{tabular}\\
        $^{a}$Effective wavelength retrieved from \url{http://svo2.cab.inta-csic.es/theory/fps3/index.php?mode=browse}.\\
        $^{b}$Central wavelength of the spectral windows used for the continuum image (see Fig. \ref{fig1}e).\\
        $^{c}$3$\sigma$ upper limit, which is not taken into account in the SED fitting analysis.\\ 
        $^{d}$Indicate that the upper limits are too large values to be shown in the Fig. \ref{fig2}.
        
        \end{flushleft}
    \end{table*}
    \begin{table*}
    	\centering
    	\caption{
    	Multi-wavelength data of GRB 050915A host galaxy.
        }
    	\label{tab7}
    	\begin{flushleft}
    	\begin{tabular}{|l|c|c|c|c|}\hline
    	\multicolumn{5}{|c|}{GRB050915A}  \\ \hline \relax
        Observed wavelength ($\mu$m) & Band & Flux density ($\mu$Jy) & Telescope/Instrument & Reference \\ \hline \relax
    
        0.47068$^{a}$& g & 0.33$\pm$0.06& Keck-I/LRIS & \citet{Perley2013a} \\
        0.54037$^{a}$&V& 0.38$\pm$0.04& Keck-I/LRIS & \citet{Perley2013a}\\ 
        0.6550$^{a}$&R& 0.49$\pm$0.08& VLT/FORS2 & \citet{Hjorth2012};\citet{Perley2013a}\\ 
        0.75497$^{a}$&I& 0.73$\pm$0.06& Keck-I/LRIS & \citet{Perley2013a}\\ 
        2.1600&K&3.57$\pm$0.87& VLT/ISAAC & \citet{Hjorth2012};\citet{Perley2013a} \\
        3.5075$^{a}$&3.6& 7.86$\pm$0.76&Spitzer/IRAC & \citet{Perley2013a} \\ 
        4.4365$^{a}$&4.5& 7.97$\pm$0.77&Spitzer/IRAC & \citet{Perley2013a} \\ 
        873$^{b}$&band7&308$\pm$87&ALMA & This work \\
        33300&4-cm band&<37$^{c}$&ATCA & \citet{Michalowski2012} \\
        54500&4-cm band&<44$^{c}$&ATCA &  \citet{Michalowski2012}\\
        216000&16-cm band&<59$^{c}$&ATCA & \citet{Michalowski2012}\\
        \hline 
        \end{tabular}\\
        $^{a}$Effective wavelength retrieved from \url{http://svo2.cab.inta-csic.es/theory/fps3/index.php?mode=browse}.\\
        $^{b}$Central wavelength of the spectral windows used for the continuum image (see Fig. \ref{fig1}f).\\
        $^{c}$3$\sigma$ upper limit, which is not taken into account in the SED fitting analysis.
        \end{flushleft}
    \end{table*}
    
    \section{SED Modelling and results}
    \label{analysis_results}
    \texttt{CIGALE}\footnote{\label{fn1}\url{https://cigale.lam.fr/about/}}
    \citep{Burgarella2005,Noll2009,Boquien2019} was designed to study the evolution of galaxies. \texttt{CIGALE} allowed us to investigate the physical properties such as SFR, stellar mass, IR luminosity, and dust mass of galaxies by comparing the SEDs of modelled galaxies to observed galaxies.
    It provides a large number of different models with different star-formation histories, AGNs, nebular emissions and dust emissions.
    
\begin{figure*}
    
        \includegraphics[width=2.2in]{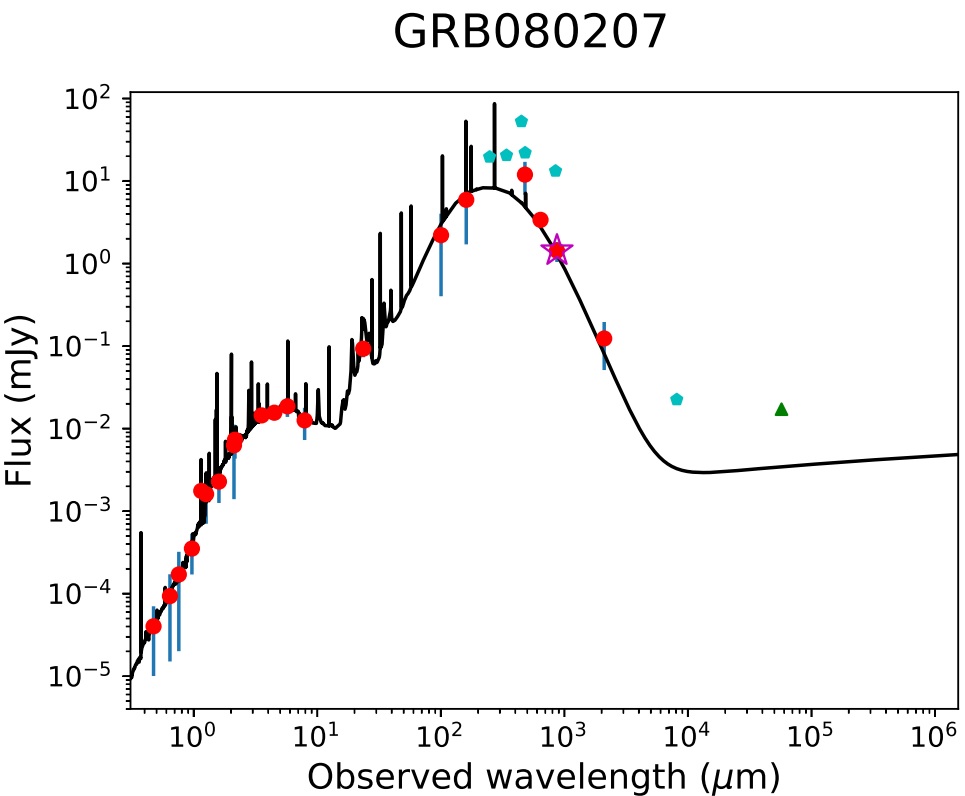}
    	\includegraphics[width=2.2in]{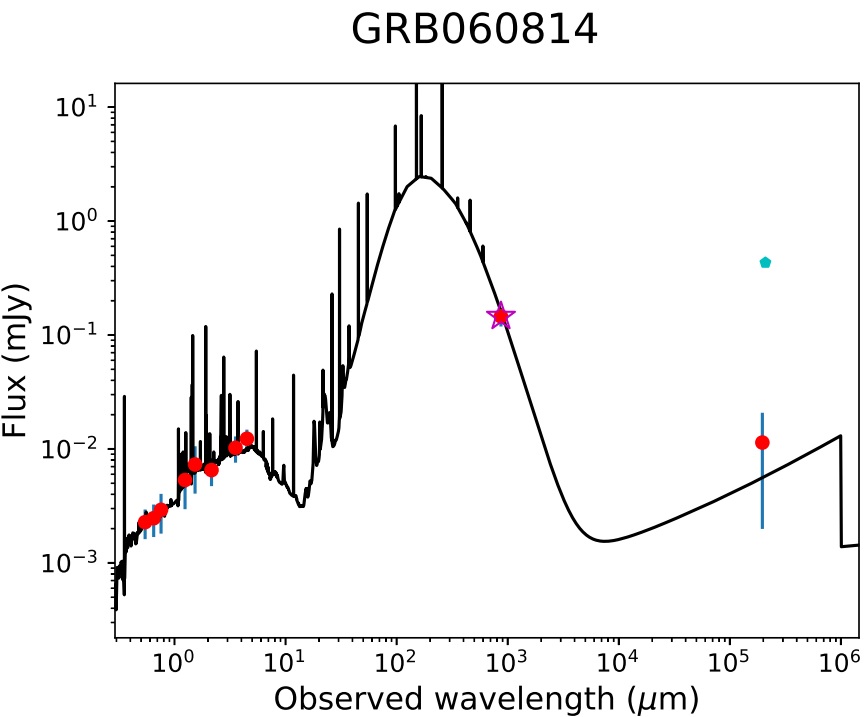}
    	\includegraphics[width=2.2in]{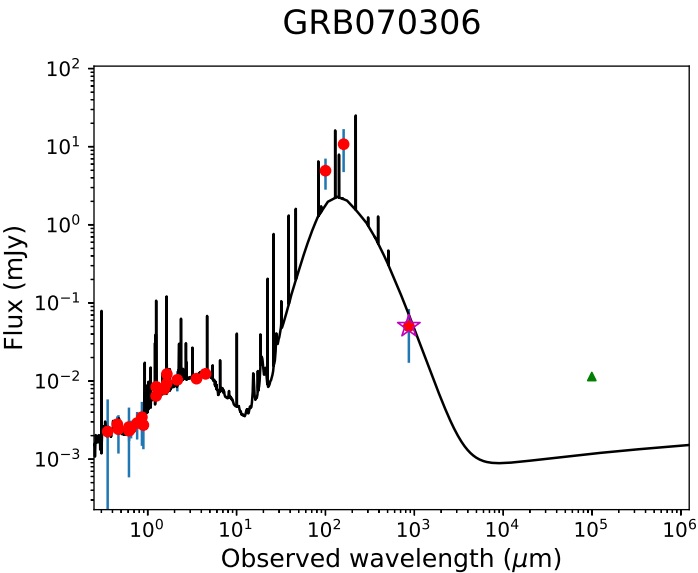}
        \includegraphics[width=2.2in]{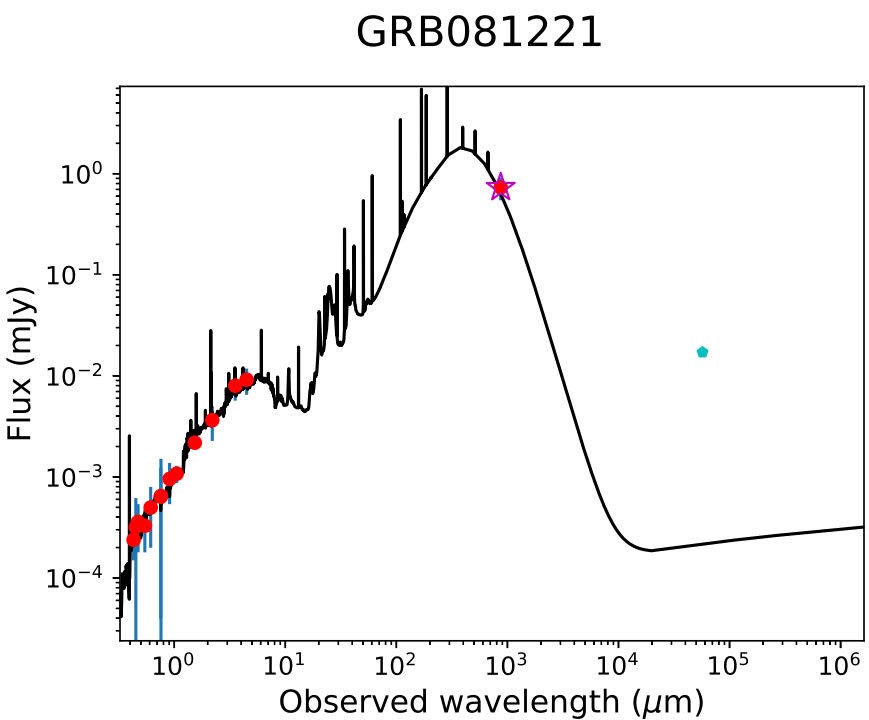}
    	\includegraphics[width=2.2in]{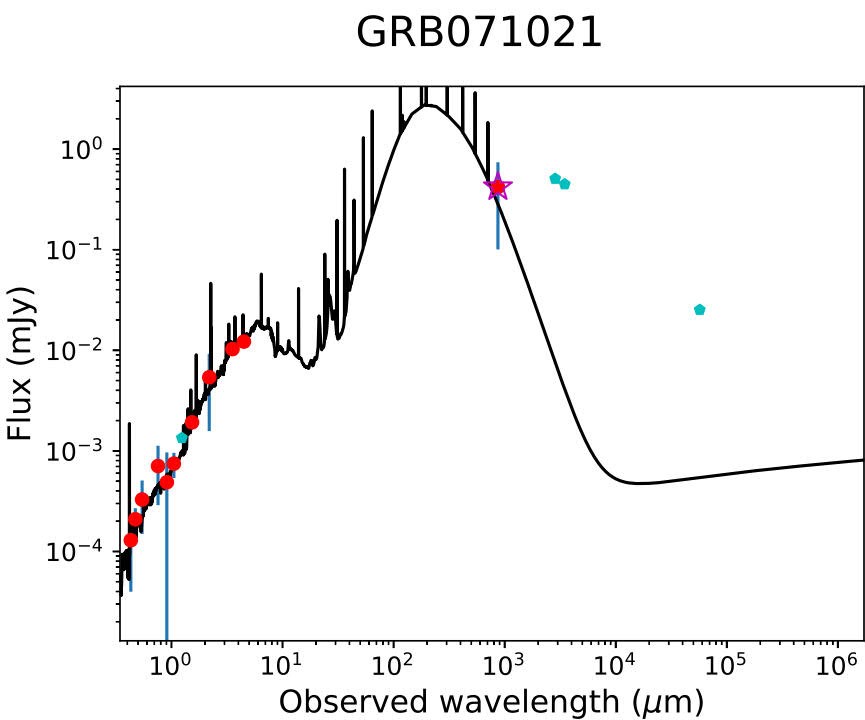}
    	\includegraphics[width=2.2in]{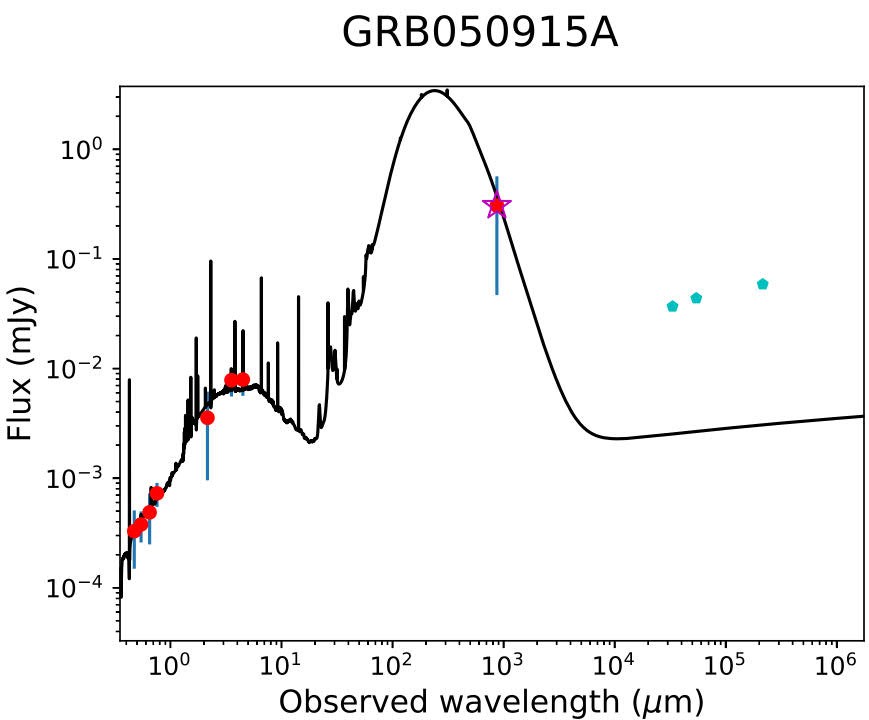}
    	\includegraphics[width=2.2in]{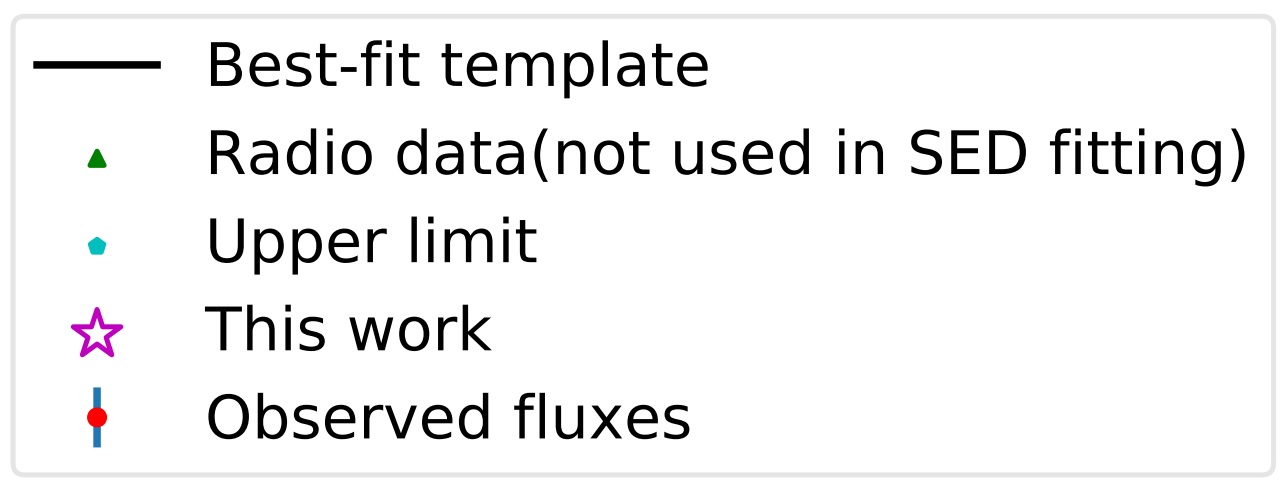}
       \caption{
        Spectral energy distributions of 
        six host galaxies.
        Among four different configurations of SED fittings, a result from the best configuration is shown for each host galaxy.
        The best-fit templates are displayed as solid lines.
        Red points are observed data used in the SED fitting analysis.
        Green triangles are observed radio data that are excluded in the SED fitting analysis.
        Blue pentagons are upper limits reported in previous studies.
        Magenta stars show ALMA data in this work.}

        \label{fig2}
    \end{figure*}

    We used \texttt{CIGALE} (version: 2018.0) to perform the SED fitting. The best fit models are shown in Fig. \ref{fig2}. 
    In this paper, we chose a single exponential star formation history (sfh2exp hereafter) and delayed SFH with optional exponential burst (sfhdelayed hereafter) to simulate star-formation history (SFH) model, 
    \citet{Bruzual2003} (hereafter bc03) to model the single stellar populations (SSP), `nebular' for nebular emission model, `dustatt\_powerlaw' for dust attenuation model, and `redshifting' for redshift+IGM model. We used the module from  \citet{Dale2014} for AGN and dust emission models except for GRB050915A because the reduced $\chi^2$ was too large for this host when this model was applied. 
    For GRB050915A, we selected the model by  \citet{Draine2014} for dust emission and \citet{Fritz2006} for the AGN component. The module we used and the parameters setting are shown in Tables \ref{tab8} and \ref{tab9}.
    
    The SFRH of sfh2exp is assumed in \texttt{CIGALE} as follows:
    \begin{equation}
    \label{eq1}
    \rm SFR( \it t)\propto \left\{
    \begin{aligned}
    \rm exp(\it -t/\tau_{0}) & \quad \rm if \quad \it t<t_{0}-t_{1}\\
    \rm exp(\it -t/\tau_{0})+k\times \rm exp(\it -t/\tau_{1}) & \quad \rm if \quad \it t\ge t_{0}-t_{1} \\
    \end{aligned}
    \right.
    \end{equation}
    where
    $t_0$ is the time when the galaxy started forming stars and $t_1$ is the time when the burst of star formation occurs. 
    $k$ is the relative amplitude of the second exponential. $\tau_0$ and $\tau_1$ represent the e$-$folding times of the main stellar populations and the late starburst population model, respectively \citep{Boquien2019}.
    
    We also assumed delayed SFH in 
    \texttt{CIGALE}:
    \begin{equation}
    \label{eq2}
    \rm SFR( \it t)\propto t/\tau^2 \times {\rm exp}(-t/\tau) \quad {\rm for} \quad 0\le t\le t_o, 
    \end{equation}
    where $t_o$ means the onset of star formation and $\tau$ 
    is the time when SFR peaks \citep{Boquien2019}.
    
    Furthermore, stellar spectra were computed by \texttt{CIGALE} from both SFH model and SSP model. SSP($\lambda$,t) were calculated at the age $t$ and wavelength $\lambda$ as the spectrum of a single population. The spectrum S($\lambda$) of the object of interest is $\sum_{t} {\rm SSP}(\lambda,t)\times {\rm SFH}(t)$, where ${\rm SFH}(t)$ is the star-formation rate at age $t$. 
    
    We used nebular modules,
    following
    \citet{Inoue2010} and \citet{Inoue2011}, to deal with the emission line template.
    Based on the model from \citet{Charlot2000}, `dustatt\_powerlaw' module provides a single power law for both young and old stars.
    
    \texttt{CIGALE} provided three types of modules to model dust emission. One of the newest is \citet{Dale2014}, which can adjust the parameters such as AGN fraction and alpha slope. Based on \citet{Dale2002}, the module \citet{Dale2014} in \texttt{CIGALE} presents the dust templates which are based on a sample of adjacent star-forming galaxies.
    The other option of dust emission we also selected is \citet{Draine2014} (dl2014 in \texttt{CIGALE}). This model is refined in detail from 
    \citet{Draine2007}. This model tries to produce a galaxy with extinction curve by considering a mixture of amorphous silicate and graphitic grains in the galaxy.
    
    The AGN model we applied for GRB050915A is \citet{Fritz2006}. This model analyzes the thermal dust emission and scattered emission by dust in the toroidal structure surrounding the AGN.
    
    We also used the redshifting module by \texttt{CIGALE}. Based on the previous study \citep{Meiksin2006}, redshifting module not only redshifts the spectrum by multiplying the wavelengths by $1+z$ and dividing the spectrum with $z+1$, but also considers the absorption of the radiation in short wavelengths by the intergalactic medium. 
    
    We classified our GRB host galaxies into two groups: Group A with radio detection (GRB080207, GRB060814 and GRB070306) and Group B without radio detection (GRB081221, GRB071021 and GRB050915A). For all host galaxies, we selected two IMFs to model the best-fit spectrum: \citet{Salpeter1955} and \citet{Chabrier2003}. We also used the four configurations of the model setting, with and without AGN component, and then with and without the radio data and radio emission for Group A. For Group B, we tested the condition with and without AGN component for best template. The results under these conditions with reduced $\chi^{2}$ are shown in Table \ref{tab10} for Group A and Table \ref{tab11} for Group B.
    
    Based on the reduced $\chi^2$, we selected the best-fit galaxy template from the conditions mentioned in the previous paragraph. However, we caution readers that our terminology of ``best" is in the sense that in some cases there exist other templates with similar 
    reduced $\chi^2$ values. Therefore, in such a case, we do not strongly exclude physical properties derived from other configurations in the SED fittings. The photometry including the ALMA data were fitted successfully with reduced $\chi^2$ of 2.32, 1.00, 2.34, 0.94, 0.99 and 0.97 for GRB080207, GRB060814, GRB070306, GRB081221, GRB071021 and GRB050915A host galaxies, respectively.
    
    We reproduced the observed SEDs with \texttt{CIGALE} and obtain the physical properties such as SFR, dust luminosity, and stellar mass.
    The derived physical properties are compared with previous works \citep[e.g.,][]{Hunt2014, Hashimoto2019,Perley2013a,Perley2015, Kruhler2011} in Table \ref{tab12} for Group A and Table \ref{tab13} for Group B. 
    
    
    
    

        \begin{table*}
    	\centering
    	\caption{
    	Parameter setting in the SED fitting with \texttt{CIGALE} for Group A
        }
    	\label{tab8}
    	\begin{flushleft}
    	\begin{tabular}{|c|c|c|c|}\hline
    	 & {GRB080207} & {GRB060814} & {GRB070306} \\ \hline \relax
    	Parameter & \multicolumn{3}{|c|}{Value}\\ \hline \relax
    	 & \multicolumn{3}{|c|}{SFH(sfh2exp)} \\ \hline \relax
    	$\tau_{\rm main}$ & 2,4,5,8,10 & 250, 500, 1000, 2000,	4000, 6000, 8000 & 5000, 7000, 10000, 12000\\ \relax
    	$\tau_{\rm burst}$ & 25,50,75 & \multicolumn{2}{|c|}{50} \\ \relax
    	$f_{\rm burst}$ & \multicolumn{3}{|c|}{0.01}\\ \relax
    	age & 2100,2125,2150 & 250,	500, 1000, 2000, 4000, 8000, 10000, 12000 & 1000, 2000, 3000, 5000 \\  \relax
    	${\rm burst}_{\rm age}$ & \multicolumn{3}{|c|}{20}\\ \relax
    	${\rm SFR}_{0}$ & \multicolumn{3}{|c|}{1.0}\\ \hline \hline \relax
	
    	 & \multicolumn{3}{|c|}{SSP\citep{Bruzual2003}} \\ \hline \relax
    	IMF & \multicolumn{3}{|c|}{\citet{Salpeter1955} and \citet{Chabrier2003}}\\ \relax
    	Metallicity & \multicolumn{3}{|c|}{0.02} \\ \relax
    	${\rm separation}_{\rm age}$ & \multicolumn{3}{|c|}{10}\\ \hline \hline \relax

    	 & \multicolumn{3}{|c|}{Nebular} \\ \hline \relax
    	logU & \multicolumn{3}{|c|}{-2.0}\\ \relax
    	$f_{\rm esc}$ & \multicolumn{3}{|c|}{0.0} \\ \relax
    	${f}_{\rm dust}$ & \multicolumn{3}{|c|}{0.0}\\ \relax
    	${\rm lines}_{\rm width}$ & \multicolumn{3}{|c|}{0.0}\\
    	\hline \hline \relax
    	
    	 & \multicolumn{3}{|c|}{Dust attenuation} \\ \hline  \relax
    	 $\rm Av_{young}$ & \multicolumn{3}{|c|}{0.0, 0.5, 1.0, 1.5, 2.0, 2.5, 3.0}\\ \relax
    	 ${\rm Av\_old\_factor}$ & \multicolumn{3}{|c|}{0.44}\\ \relax
    	 uv\_bump\_wavelength & \multicolumn{3}{|c|}{217.5}\\ \relax
    	 uv\_bump\_width & \multicolumn{3}{|c|}{35.0}\\ \relax
    	 uv\_bump\_amplitude & \multicolumn{2}{|c|}{0.0, 1.0, 2.0, 3.0} & 0.0, 0.5, 1.0, 1.5, 2.0, 2.5, 3.0\\ \relax
    	 power\_slope & \multicolumn{2}{|c|}{-0.7} & -0.4, -0.7\\
    	 \hline \hline \relax
    	 & \multicolumn{3}{|c|}{Dust emission\citep{Dale2014}} \\ \hline \relax
    	 AGN fraction$^{a}$ ($\times 10000$) & 1, 5, 8, 10, 15, 20, 30 & 10, 20, 50, 100, 150, 200, 500, 1000 & 10, 20, 50, 100, 200, 500, 1000\\ \relax
         $\alpha$ slope & 1.0, 1.25, 1.5, 1.75, 2.0 & 1.0, 1.5, 2.0, 2.5 & 0.5, 1.0, 2.0, 3.0, 4.0\\ \hline \hline \relax
    	 & \multicolumn{3}{|c|}{Radio$^{b}$} \\ \hline \relax
    	 $\alpha_{\rm radio}$ & \multicolumn{3}{|c|}{0.6, 0.7, 0.8, 0.9, 1.0} \\ \relax
    	 qir & \multicolumn{3}{|c|}{2.58} \\ \hline \hline \relax
        \end{tabular}\\
        {\bf Notes.} Detailed definition of these parameters are described in Appendix C of \citet{Boquien2019} \\ \relax
        $^{a}$In only radio emission condition and none of those two, AGN fraction set to 0.\\ \relax
        $^{b}$In only AGN condition and none of those two, we did not use radio module. 
        
        \end{flushleft}
    \end{table*}
    
    \begin{table*}
    	\centering
    	\caption{
    	Parameter setting in the SED fitting with \texttt{CIGALE} for Group B
        }
    	\label{tab9}
    	\begin{flushleft}
    	\begin{tabular}{|c|c|c|c|}\hline
    	 & {GRB081221} & {GRB071021} & {GRB050915A} \\ \hline \relax
    	Parameter & \multicolumn{3}{|c|}{Value}\\ \hline \relax
    	 SFH& sfhdelayed &\multicolumn{2}{|c|}{sfh2exp} \\ \hline \relax
    	$\tau_{\rm main}$ & 1000,1500,2000,2500&\multicolumn{2}{|c|}{250,400,500,750,1000}\\ \relax
    	$\tau_{\rm burst}$ & \multicolumn{3}{|c|}{50} \\ \relax
    	$f_{\rm burst}$ & 0.0&\multicolumn{2}{|c|}{0.01}\\ \relax
    	age &1000,1500,2000,2500,3000,3500,4000&  250, 500, 1000, 1500, 2000, 2500, 4000 &250,500,750,100,1250,1500  \\ \relax
    	${\rm burst}_{\rm age}$ & \multicolumn{3}{|c|}{20}\\ \relax
    	${\rm SFR}_{0}$ & \multicolumn{3}{|c|}{1.0}\\ \hline \hline \relax
    	 & \multicolumn{3}{|c|}{SSP\citep{Bruzual2003}} \\ \hline \relax
    	IMF & \multicolumn{3}{|c|}{\citet{Salpeter1955} and \citet{Chabrier2003} }\\ \relax
    	Metallicity & \multicolumn{2}{|c|}{0.05}&0.0001 \\ \relax
    	${\rm separation}_{\rm age}$ & \multicolumn{3}{|c|}{10}\\ \hline \hline \relax
    	
    	& \multicolumn{3}{|c|}{Nebular} \\ \hline \relax
    	logU & \multicolumn{3}{|c|}{-2.0}\\ \relax
    	$f_{\rm esc}$ & \multicolumn{3}{|c|}{0.0} \\ \relax
    	${f}_{\rm dust}$ & \multicolumn{3}{|c|}{0.0}\\ \relax
    	${\rm lines}_{\rm width}$ & \multicolumn{3}{|c|}{0.0}\\  \hline \hline \relax
    	 & \multicolumn{3}{|c|}{Dust attenuation} \\ \hline \relax
    	 $\rm Av_{young}$ & \multicolumn{3}{|c|}{0.0, 0.5, 1.0, 1.5, 2.0, 2.5, 3.0}\\ \relax
    	 ${\rm Av\_old\_factor}$ & \multicolumn{3}{|c|}{0.44}\\ \relax
    	 uv\_bump\_wavelength & \multicolumn{3}{|c|}{217.5}\\ \relax
    	 uv\_bump\_width & \multicolumn{3}{|c|}{35.0}\\ \relax
    	 uv\_bump\_amplitude & \multicolumn{3}{|c|}{ 0.0, 1.0, 2.0, 3.0} \\ \relax
    	 power\_slope & \multicolumn{3}{|c|}{-0.7}\\
    	 \hline \hline \relax
    	 & \multicolumn{2}{|c|}{Dust emission\citep{Dale2014}}& \citet{Draine2014}\\ \hline \relax
    	 AGN fraction$^{a}$ ($\times 10000$) &9,10,11,50,100,1000&50,100,200,500&none\\ \relax
     	 $\alpha$ slope&1.8750,1.9375,2.0,2.0625,2.1250,2.1875,2.25&1.25,1.375,1.5,1.625,1.75,2,2.25&none \\ \relax
    	 qpah &\multicolumn{2}{|c|}{none}&0.47,1.12,1.77,2.5\\  \relax
    	 umin &\multicolumn{2}{|c|}{none}&0.1,1,2,3,4,5,10,15,20,25,30\\  \relax
    	 $\alpha$ &\multicolumn{2}{|c|}{none}&2.2,2.3,2.4,2.5,2.6,2.7,2.8\\ \relax
    	 $\gamma$ &\multicolumn{2}{|c|}{none}&0.1,0.2,0.3,0.4\\ \hline \hline \relax
    	 & \multicolumn{3}{|c|}{AGN$^{a}$\citep{Fritz2006}} \\ \hline \relax
    	 \rm AGN fraction ($\times 10000$)&\multicolumn{2}{|c|}{none}&10,90,100,300,350,400,450,500\\ \hline \hline \relax
    	 
        \end{tabular}\\

    	 {\bf Notes.} Detailed definition of these parameters are described in Appendix C of \citet{Boquien2019} \\ \relax
        $^{a}$This module is only present for the AGN configuration
       
       \end{flushleft}
    \end{table*}
    
    \section{Discussion}
    \label{discussion}
    
    We discuss the best-fit spectra in different model settings.
    The configurations contain on-off options of AGN and radio components for Group A. 
    For Group B, we discuss the results with and without the AGN component.
    Here, we consider the AGN component because some GRB hosts actually indicate hints of AGNs \citep[e.g.,][]{Hatsukade2019}.
    The options to include and exclude radio data are due to a possible contamination from a long-lived radio afterglow in the radio flux of the host galaxy \citep[e.g., ][]{Perley2013b}.
    The physical parameters and results are shown in Tables \ref{tab10} and \ref{tab11} for each GRB host galaxy. In addition, we also compare our results with previous papers in Tables \ref{tab12} and \ref{tab13}.
    
    GRB host galaxies could have different stellar IMFs compared to a normal galaxy  \citep[e.g.,][]{Hashimoto2019}. As a result, it is important to model the IMF for each host galaxy. 
    For this purpose, we tested different IMFs from \citet{Salpeter1955} and \citet{Chabrier2003}. We simply decide the better fit by comparing the reduced $\chi^2$ between these two different IMFs. 
    We caution readers that our testing scheme is very simple, and therefore, the IMF differences could be masked by the simplifications of the model.

    For fair comparisons, we converted the derived physical properties of those with the same IMF in previous studies by the factor from \citet{Madau2014}. The differences are shown in Tables \ref{tab12} and \ref{tab13} after the conversion of IMF.

    \subsection{Group A (Radio detected hosts)}
    \subsubsection{GRB080207 host galaxy}
    We compare reduced $\chi^2$ derived from different configurations of the SED fitting for the GRB 080207 host.
    We found that the best-fit SED contains an AGN component excluding the radio emission with the smallest reduced $\chi^2$ value of 2.32.
    The flux upper limits reported in previous studies are consistent with the best-fit template (Fig. \ref{fig2}).
    The AGN fraction of the best-fit model is $0.0012\pm0.0007$.
    Comparing with the fitting of model which does not contain the AGN component, the best-fit template is better in the short wavelengths of observed data.
    If the radio data and model of radio emission are included in the fitting procedure, the reduced $\chi^2$ becomes larger. One possible reason is that a long-lived radio afterglow has contaminated the radio flux \citep[e.g., ][]{Perley2013b}.
    The SFR of the best-fit template including the AGN component is $397\pm20 M_{\odot}$ yr$^{-1}$, which is slightly lower than SFRs of other models without AGN component (Table \ref{tab10}).
    
    To check the SFRs derived by SED fittings, we use the empirical formula between IR luminosity and SFR in \citet{Kennicutt1998b}:
    \begin{equation}
    \label{eq3}
    {\rm SFR_{Kennicutt}} (M_{\odot} {\rm yr}^{-1})=4.5\times10^{-44}L_{\rm IR}({\rm ergs}^{-1}).
    \end{equation}
    %
    This equation is equivalent to
    \begin{equation}
    \label{Kennicutt1998}
    {\rm SFR_{Kennicutt}}(M_{\odot} {\rm yr}^{-1})=1.722\times10^{-10}L_{\rm IR}(L_{\odot}).
    \end{equation}
    The IR luminosity in \citet{Kennicutt1998b} is defined as the luminosity integrated from 8 to 1000 $\mu$m.
    We defined the IR luminosity of our results as the luminosity emitted by dust, 
    which is almost similar to $L_{IR}$ in most of the star-forming galaxies.
    The SFRs derived from Eq. \ref{Kennicutt1998} are 305, 313, and 310 $M_{\odot}$ yr$^{-1}$ for the models including AGN, radio emission, and both, respectively.
    The SFR of the best-fit template derived by \texttt{CIGALE}, $397\pm20 M_{\odot}$ yr$^{-1}$, is closer to these values than that of other templates in Table \ref{tab10}.
    
    In Table \ref{tab10}, the Salpeter IMF \citep{Salpeter1955} fits the observed SED better for all of the four model configurations.
    The derived IR luminosities and stellar masses are consistent between all of the model settings.
    
    We also demonstrate the fitting and results from \citet{Hunt2014} and \citet{Hashimoto2019} in Table \ref{tab12}. 
    In order to compare with previous works, we scale our estimations of SFR and stellar mass of GRB080207 host galaxy to \citet{Chabrier2003} IMF by using a conversion factor in \citet{Madau2014}. 
    After comparing our stellar mass value with previous works, we conclude that our stellar mass is closer to the estimation in \citet{Hashimoto2019}. 

    Interestingly, all of the four different conditions suggest that this host galaxy is an ULIRG because its IR luminosity is larger than $10^{12}L_{\odot}$, which is consistent with both previous papers.
    
    However, the SFR derived by the SED fitting in this work is much higher ($\sim 50\%$ to $\sim 100\% $) than that in both \citet{Hunt2014} and \citet{Hashimoto2019}. 
    The IR luminosity of the best-fit template corresponds to $\rm{SFR}_{Kennicutt}=305(M_{\odot} yr^{-1})$ assuming Kennicutt's law (Eq. \ref{Kennicutt1998}).
    This value is much closer to the SFR derived from the SED fitting analysis in this work than that in previous papers.
    Our new SFR measurement is more reasonable than that in previous studies if GRB hosts follows the Kennicutt's law. The reason for the difference is due to the new ALMA data.
    The ALMA data allowed us to cover the wavelength around the peak of the SED of dust thermal emission (Fig. \ref{fig2}), which results in a more reliable measurement of SFR.
    
    \subsubsection{GRB060814 host galaxy}
    We present physical parameters of the best-fit SED template of the GRB060814 host galaxy in Table \ref{tab10}.
    A comparison between this work and previous works is summarised in Table \ref{tab12}.
    The physical parameters of the best-fit template are ${\rm SFR_{SED}}=56\pm9M_{\odot}$ yr$^{-1}$ and $M_{*}=(2.68\pm0.76)\times10^{10}M_{\odot}$. 
    The reduced $\chi^2$ is 1.00 which is the closest to 1 in our four conditions.
    The flux upper limits reported in previous studies are consistent with the best-fit template (Fig. \ref{fig2}).
    The models fit the observed SED better with the \citet{Chabrier2003} than \citet{Salpeter1955}, except for the model including both AGN and radio components.
    If the radio emission is not included in the SED-fitting, the reduced $\chi^2$ becomes smaller than 1.0, suggesting over-fitting.
    Including both AGN component and radio emission also seems to cause the over-fitting with a reduced $\chi^{2}$ of 0.53.
    Only the model with radio emission without AGN component provides the best-fit template with a reasonable reduced $\chi^{2}$.
    In general, an ideal fitting should correspond to a reduced $\chi^2$ of 1.0 if all the errors are estimated perfectly.
    However this is probably not the case in our analysis.
    Therefore, we demonstrate all the results derived from the four different model configurations in Table \ref{tab10}.
    
    We also estimated SFRs of the host galaxy from the IR luminosity and Kennicutt's law in the same manner as the GRB 080207 host galaxy.
    The estimated SFRs are 54, 69, and 66 $M_{\odot}$ yr$^{-1}$ for model configurations including AGN, radio, and both, respectively.
    
    We also compared the parameters derived from our best fit and from previous works.
    In this work, the stellar mass of the best-fit template is $M_{*}=(2.68\pm0.76) \times 10^{10} M_{\odot}$.
    This value is consistent with $M_{*}=1.6^{+1.4}_{-0.6}\times 10^{10} M_{\odot}$ in \citet{Perley2015}, which is an updated value compared to 
    \citet{Perley2013a}.
    The SFR derived from the best-fit template is $56\pm9 M_{\odot}$ yr$^{-1}$, 
    which
    is much lower than ${\rm SFR_{SED}}=238.2^{+49.6}_{-24.0}$ and $209^{+27}_{-53} M_{\odot}$ yr$^{-1}$ 
    from 
    \citet{Perley2013a,Perley2015}.
    Since the previous works lack the rest-frame FIR continuum data, they probably overestimated the SFR.
    The ALMA detection in the rest-frame FIR of the host galaxy enabled us to calculate the SFR more accurately.
    The SFR derived from the SED fitting in \texttt{CIGALE} (${\rm SFR_{SED}}=56\pm9 M_{\odot}$ yr$^{-1}$) is closer to the estimation using Kennicutt's law (SFR$_{\rm Kennicutt}=69 M_{\odot}$ yr$^{-1}$) compared to previous studies.
    
    \subsubsection{GRB070306 host galaxy}
    The GRB070306 host galaxy is faint in the rest-frame FIR and marginally detected with ALMA. 
    We performed SED fittings for the GRB070306 host galaxy. 
    The physical properties are shown in Table \ref{tab10} including parameters from different model settings. 
    In all cases, the \citet{Chabrier2003} IMF performs better than the \citet{Salpeter1955} IMF.
    The reduced $\chi^2$ values are almost the same for all configurations, but the smallest reduced $\chi^{2}$ value was attained in the model without AGN and radio component, which is 2.324.
    Physical parameters of the best-fit template are ${\rm SFR_{SED}}=38\pm2 M_{\odot}$ yr$^{-1}$, $M_{*}=(3.05\pm0.44)\times 10^{10} M_{\odot}$, and $L_{\rm IR}=(2.59\pm0.26)\times 10^{11} L_{\odot}$.
    The stellar masses, SFRs and IR luminosities derived from all the four conditions are consistent within errors. 
    The derived IR luminosities for the four settings correspond to SFR$_{\rm Kennicutt}=44$ to $46 M_{\odot}$ yr$^{-1}$ assuming the Kennicutt's law.
    The best-fit template indicates SFR$_{\rm Kennicutt}=45 M_{\odot}$ yr$^{-1}$.
    This value is similar to the ${\rm SFR_{SED}}=38\pm2 M_{\odot}$ yr$^{-1}$ derived from our SED fitting analysis in contrast to ${\rm SFR_{SED}}=12.59^{+12.53}_{-4.65}, 144.1$, and $17^{+7}_{-5} M_{\odot}$ yr$^{-1}$ in previous works \citep{Kruhler2011,Hunt2014,Perley2015}.
    The SFR$_{\rm SED}=38 M_{\odot}$ yr$^{-1}$ is slightly larger than that in \citet{Perley2015}, i.e., $SFR={\rm SFR_{SED}}17^{+7}_{-5} M_{\odot}$ yr$^{-1}$.
    The difference of SFR is probably due to the new ALMA data in this work.
    
    
    \subsection{Group B (Radio non-detected hosts)
    }
    For Group B, due to the non-detection of radio in previous studies, we did not include the radio emission in the procedure of SED fitting.
    We discuss the comparison of the remaining three GRB hosts: GRB081221, GRRB071021 and GRB050915A in Table \ref{tab13}. 
    These three hosts have similar redshifts and optical faintnesses.
    They show higher stellar masses and SFRs in previous works \citep{Michalowski2012,Perley2013a,Hunt2014}.
    
    \subsubsection{GRB081221 host galaxy}

    Our results show that, if we consider the AGN component in our fitting process, the reduced $\chi^2$ becomes smaller, suggesting over-fitting. 
    An ideal fitting should correspond to a reduced $\chi^2$ of 1.0 if all the errors are estimated perfectly. 
    As a result, we show the results for the configuration with AGN component in Table \ref{tab11}.

 Our results show that \citet{Salpeter1955} IMF is better. In order to compare with previous work \citep{Perley2013a}, we convert our stellar mass and SFR value from \citet{Salpeter1955} to \citet{Chabrier2003} by the factor from \citet{Madau2014}. The stellar mass is consistent with that in \citet{Perley2013a} within the uncertainties. However, for the SFR values, 
 our value is lower than previous work. When we transform our $L_{\rm IR}$ to SFR through \citet{Kennicutt1998b},
 we obtained approximately $\rm{SFR}_{Kennicutt}=52\pm3(M_{\odot}yr^{-1})$ which is very close to our estimation ${\rm SFR_{SED}}=56(M_{\odot}yr^{-1})$. 
 The reason for the difference between our work and previous works is probably that no FIR detection was included in previous works, causing their SFR to be overestimated.
 Therefore, the new FIR detection from ALMA gives us more accurate SFR values if the host galaxy follows the Kennicutt law.
    
    \subsubsection{GRB071021 host galaxy}
    Based on the reduced $\chi^2$, if we add the AGN component to this modelling, the resulting fit is also an over-fit.
    The reduced $\chi^2$ with AGN is 0.68 and without AGN is 0.99. The results of two conditions with and without AGN component are summarised in Table \ref{tab11}. 
    The IMF configuration 
    of \citet{Chabrier2003} 
    performed better in 
    the two conditions.

 Here we compare our results involving stellar mass, SFR, and dust luminosity to those in \citet{Hunt2014} and \citet{Perley2013a} in Table \ref{tab13}. 
 The stellar mass is consistent within the uncertainty between \citet{Perley2013a} and ours. The SFR values in the three previous works differ from each other, and comparing them with our result shows that our value ${\rm SFR_{SED}}=90\pm5(M_{\odot}yr^{-1})$ is the lowest. The SFR transformation of IR luminosity by \citet{Kennicutt1998b} indicates $\rm{SFR}_{Kennicutt}= 129\pm28(M_{\odot}yr^{-1})$ which is very close to our results compared with previous studies. 
  Moreover, there was no FIR photometry included in previous works, which is probably the cause of over-estimation of SFR values.
 
    
    \subsubsection{GRB050915A host galaxy}
    
    In this host galaxy, we found several candidates for the host galaxy (Fig. \ref{fig1}f). We chose the closest one between the candidate and the host position. The separation between the two is $0.17$ arcsec.
   
    We performed again the setting for two configurations: with and without AGN component. The results show 
    an over-fit when we added the AGN component (reduced $\chi^2$ = 0.47). However, the reduced $\chi^2$ value is $0.97$ when we removed the AGN component. Physical parameters revealed by SED fitting using these two configurations are shown in Table \ref{tab11} with the better IMF configuration by \citet{Chabrier2003}.
    
    There are significant differences in stellar masses and SFRs between this work and 
    \citet{Perley2013a}.
    Also, the transformation by using \citet{Kennicutt1998b} shows that $\rm{SFR}_{Kennicutt}=148\pm7(M_{\odot}yr^{-1})$.
    This value is slightly larger compared to all the previous results and our results. In this work, the new photometry of the host galaxy with ALMA and previous upper limits are consistent with the best-fit template. The derived SFR including the rest-frame FIR data is likely to be more reliable than that in previous studies.
    
    
    \begin{table*}
    	\centering
    	\caption{
    	Physical parameters of GRB hosts for Group A in different model conditions.
        }
    	\label{tab10}
    	\begin{flushleft}
    	\begin{tabular}{|c|c|c|c|c|}\hline
    	\multicolumn{5}{|c|}{GRB 080207}  \\ \hline \relax
    	Physical parameters & AGN$^{a}$ &Radio emission& Both & None of these two\\ \hline \relax
        Stellar mass ($M_{\odot}$) &  $(3.25\pm{0.16})\times10^{11}$ & $(3.32\pm{0.17})\times10^{11}$ & $(3.29\pm{0.17})\times10^{11}$ & $(3.29\pm{0.17})\times10^{11}$ \\ \relax
        SFR ($M_{\odot}$ yr$^{-1}$) & $397\pm{20}$ & $409\pm{20}$ & $405\pm{20}$ & $402\pm{20}$\\ \relax
        Total IR luminosity ($L_{\odot}$) &$(1.77\pm{0.09})\times10^{12}$ & $(1.82\pm{0.09})\times10^{12}$ & $(1.80\pm{0.09})\times10^{12}$ & $(1.80\pm{0.09})\times10^{12}$ \\ \relax
        Better IMF & \citet{Salpeter1955} & \citet{Salpeter1955} & \citet{Salpeter1955} & \citet{Salpeter1955} \\ \relax
        Reduced $\chi^2$ & 2.32 & 2.56 & 2.46 & 2.45 \\
    
        \end{tabular}\\ 
    
    	\begin{tabular}{|c|c|c|c|c|}\hline
    	\multicolumn{5}{|c|}{GRB 060814}  \\ \hline \relax
    	Physical parameters& AGN &Radio emission$^{a}$& Both & None of these two\\ \hline \relax
        Stellar mass ($M_{\odot}$) & $(3.02\pm{0.83})\times10^{10}$ & $(2.68\pm{0.76})\times10^{10}$ & $(4.44\pm{1.32})\times10^{10}$ & $(3.08\pm{0.81})\times10^{10}$\\ \relax
        SFR ($M_{\odot}$ yr$^{-1}$) & $48\pm{9}$ & $56\pm{9}$ & $85\pm{14}$ & $47\pm{10}$ \\ \relax
        Total IR luminosity ($L_{\odot}$) & $(3.15\pm{1.14})\times10^{11}$ & $(4.03\pm{1.07})\times10^{11}$ & $(3.85\pm{1.05})\times10^{11}$ &
        $(3.05\pm{1.16})\times10^{11}$ \\  \relax
        Better IMF & \citet{Chabrier2003} & \citet{Chabrier2003} & \citet{Salpeter1955} & \citet{Chabrier2003} \\ \relax
        Reduced $\chi^2$ & 0.42 & 1.00 & 0.9 & 0.53 \\ 
        \end{tabular}\\
    
    	\begin{tabular}{|c|c|c|c|c|}\hline
    	\multicolumn{5}{|c|}{GRB 070306}  \\ \hline \relax
    	Physical parameters & AGN &Radio emission& Both & None of these two$^{a}$\\ \hline \relax
        Stellar mass ($M_{\odot}$) & $(2.99\pm{0.45})\times10^{10}$ & $(3.07\pm{0.46})\times10^{10}$ & $(3.02\pm{0.47})\times10^{10}$ & $(3.05\pm{0.44})\times10^{10}$\\ \relax
        SFR ($M_{\odot}$ yr$^{-1}$)& $38\pm{2}$ & $39\pm{2}$ & $38\pm{2}$ & $38\pm{2}$\\ \relax
        Total IR luminosity ($L_{\odot}$) &  $(2.58\pm{0.26})\times10^{11}$ &  $(2.67\pm{0.20})\times10^{11}$ &  $(2.66\pm{0.22})\times10^{11}$ &  $(2.59\pm{0.26})\times10^{11}$ \\ \relax
        Better IMF & \citet{Chabrier2003} & \citet{Chabrier2003} & \citet{Chabrier2003} & \citet{Chabrier2003} \\ \relax
        Reduced $\chi^2$ & 2.34 & 2.39 & 2.40 & 2.34 \\
        \hline
        \end{tabular}\\
        $^{a}$Indicates that this model setting has the closest reduced $\chi^2$ value to 1, and further discussed in this paper.
        \end{flushleft}
    \end{table*}

    \begin{table*}
    	\centering
    	\caption{
    	Physical parameters of GRB hosts for Group B in different model conditions.
        }
    	\label{tab11}
    	\begin{flushleft}
    	\begin{tabular}{|c|c|c|}\hline
    	\multicolumn{3}{|c|}{GRB 081221}  \\ \hline \relax
    	Physical parameters$^{a}$ & AGN & No AGN$^{b}$\\ \hline \relax
        Stellar mass ($M_{\odot}$) &  $(6.54\pm{0.74})\times10^{10}$ & $(6.24\pm{0.31})\times10^{10}$\\ \relax
        SFR ($M_{\odot}$ yr$^{-1}$)& $54\pm{3}$ & $56\pm{3}$ \\ \relax
        Total IR luminosity ($L_{\odot}$) &  $(2.95\pm{0.15})\times10^{11}$ &  $(3.04\pm{+0.15})\times10^{11}$    \\ \relax
        Better IMF & \citet{Salpeter1955} & \citet{Salpeter1955} \\ \relax
        Reduced $\chi^2$ & 0.82 & 0.94 \\
        \end{tabular}\\
    
    	\begin{tabular}{|c|c|c|}\hline
    	\multicolumn{3}{|c|}{GRB 071021}  \\ \hline \relax
    	Physical parameters$^{a}$ & AGN & No AGN$^{b}$\\ \hline \relax
        Stellar mass ($M_{\odot}$) &  $(2.24\pm{0.31})\times10^{11}$ & $(1.21\pm{0.15})\times10^{11}$\\ \relax
        SFR ($M_{\odot}$ yr$^{-1}$)& $133\pm{14}$ & $90\pm{5}$ \\ \relax
        Total IR luminosity ($L_{\odot}$) &  $(6.89\pm{0.68})\times10^{11}$ &  $(7.48\pm{1.63})\times10^{11}$    \\ \relax
        Better IMF & \citet{Salpeter1955} & \citet{Chabrier2003} \\ \relax
        Reduced $\chi^2$ & 0.68 & 0.99 \\
        \end{tabular}\\
    
    	\begin{tabular}{|c|c|c|}\hline
    	\multicolumn{3}{|c|}{GRB 050915A}  \\ \hline \relax
    	Physical parameters$^{a}$ & AGN & No AGN$^{b}$\\ \hline \relax
        Stellar mass ($M_{\odot}$) &  $(9.73\pm{1.94})\times10^{10}$ & $(9.56\pm{0.48})\times10^{10}$\\ \relax
        SFR ($M_{\odot}$ yr$^{-1}$)& $77\pm{9}$ & $77\pm{4}$ \\ \relax
        Total IR luminosity ($L_{\odot}$) &  $(5.99\pm{0.62})\times10^{11}$ &  $(8.58\pm{0.43})\times10^{11}$    \\ \relax
        Better IMF & \citet{Chabrier2003} & \citet{Chabrier2003} \\ \relax
        Reduced $\chi^2$ & 0.47 & 0.97 \\
        \hline
        \end{tabular}\\
        $^{a}$Notice that this GRB did not have radio detection, so here we do not include radio emission.  \\
        $^{b}$Indicates that this model setting has the closest reduced $\chi^2$ value to 1, and further discussed in this paper.
        \end{flushleft}
    \end{table*}
    
    \begin{table*}
    	\centering
    	\caption{
    	Physical parameters of GRB hosts for Group A in comparison to previous works.
        }
    	\label{tab12}
    	\begin{flushleft}
    	\begin{tabular}{|c|c|c|c|}\hline
    	\multicolumn{4}{|c|}{GRB 080207}  \\ \hline \relax
    	Physical parameters & \citet{Hunt2014} & \citet{Hashimoto2019} & Best-fit(AGN) \\ \hline \relax
        Stellar mass ($M_{\odot}$) & 1.48$\times 10^{11}$& (1.70$^{+0.08}_{-0.35}$)$\times 10^{11}$  & $(1.98\pm{0.10})\times10^{11}$ \\ \relax
        SFR ($M_{\odot}$ yr$^{-1}$) & 170.1 & 123.40$^{+25.19}_{-21.78}$& $250\pm{13}$ \\ \relax
        Total IR luminosity ($L_{\odot}$)& 1.78 $\times 10^{12}$ & (1.82$\pm{0.23}$) $\times 10^{12}$ &$(1.77\pm{0.09})\times10^{12}$  \\ \relax
        IMF & $S2C^{1.8}$ & \citet{Chabrier2003} & $S2C^{M}$   \\ 
        \end{tabular}\\ 
    
    	\begin{tabular}{|c|c|c|c|}\hline
    	\multicolumn{4}{|c|}{GRB 060814}  \\ \hline \relax
    	Physical parameters&\citet{Perley2013a} & \citet{Perley2015} & Best-fit(Radio)\\ \hline \relax
        Stellar mass ($M_{\odot}$) &$(9.8^{+0.9}_{-0.2})\times 10^{9}$ & $(1.6^{+1.4}_{-0.6})\times 10^{10}$ & $(2.68\pm{0.76})\times10^{10}$\\ \relax
        SFR ($M_{\odot}$ yr$^{-1}$) & $238.2^{+49.6}_{-24.0}$ & $209^{+27}_{-53}$ & $56\pm{9}$ \\ \relax
        Total IR luminosity ($L_{\odot}$)& & & $(4.03\pm{1.07})\times10^{11}$ \\  \relax
        IMF & $S2C^{1.8}$ & $S2C^{1.6}$ & \citet{Chabrier2003} \\ 
        \end{tabular}\\
    
    	\begin{tabular}{|c|c|c|c|c|}\hline
    	\multicolumn{5}{|c|}{GRB 070306}  \\ \hline \relax
    	Physical parameters&\citet{Kruhler2011} & \citet{Hunt2014} & \citet{Perley2015} & Best-fit(None of those two)\\ \hline \relax
        Stellar mass ($M_{\odot}$) & $2.45\times10^{10}$ & $1.12\times10^{10}$ & $5^{+0.1}_{-0.2}\times10^{10}$ & $(3.05\pm{0.44})\times10^{10}$\\ \relax
        SFR ($M_{\odot}$ yr$^{-1}$)& $12.59^{+12.53}_{-4.65}$ & 144.1 & $17^{+7}_{-5}$ & $38\pm{2}$\\ \relax
        Total IR luminosity ($L_{\odot}$) & & $15.1\times10^{11}$ & &  $(2.59\pm{0.26})\times10^{11}$ \\ \relax
        IMF & \citet{Chabrier2003} & $S2C^{1.8}$ & $S2C^{1.6}$ & \citet{Chabrier2003} \\
        \hline
        \end{tabular}\\
        S2C means that the original setting of IMF is \citet{Salpeter1955}. But they use some factor to transform to \citet{Chabrier2003} for a fair comparison.\\
        $^{1.8}$Divided by a factor of 1.8 from \citet{Salpeter1955} to \citet{Chabrier2003}\\
        $^{1.6}$Divided by a factor of 1.6 from \citet{Salpeter1955} to \citet{Chabrier2003}\\
        $^M$Multiplied by a factor in \citet{Madau2014} (0.61 for stellar mass and 0.63 for SFR) from \citet{Salpeter1955} to \citet{Chabrier2003}.
        \end{flushleft}
    \end{table*}

    \begin{table*}
    	\centering
    	\caption{
    	Physical parameters of GRB hosts for Group B in comparison to previous works.
        }
    	\label{tab13}
    	\begin{flushleft}
    	\begin{tabular}{|c|c|c|}\hline
    	\multicolumn{3}{|c|}{GRB 081221}  \\ \hline \relax
    	Physical parameters & \citet{Perley2013a} & Best-fit(No AGN) \\ \hline \relax
        Stellar mass ($M_{\odot}$) & (3.70$^{+1.1}_{-1.2}$)$\times 10^{10}$& ($3.81\pm0.19$)$\times 10^{10}$  \\ \relax
        SFR ($M_{\odot}$ yr$^{-1}$) & 172.8$^{+22.8}_{-30.1}$ & $35\pm2$ \\ \relax
        Total IR luminosity ($L_{\odot}$)&  &$(3.04\pm{0.15})\times10^{11}$  \\ \relax
        IMF & $S2C^{1.8}$ & $S2C^{M}$   \\ 
        \end{tabular}\\ 
        

    	\begin{tabular}{|c|c|c|c|}\hline
    	\multicolumn{4}{|c|}{GRB 071021}  \\ \hline \relax
    	Physical parameters & \citet{Hunt2014}& \citet{Perley2013a} & Best-fit(No AGN) \\ \hline \relax
        Stellar mass ($M_{\odot}$) &2.51$\times 10^{11}$& (1.196$^{+0.066}_{-0.088}$)$\times 10^{11}$& (1.21$\pm{0.15}$)$\times 10^{11}$  \\ \relax
        SFR ($M_{\odot}$ yr$^{-1}$) &  288.2 &190.3$^{+25.6}_{-20.3}$ & $90\pm{5}$ \\ \relax
        Total IR luminosity ($L_{\odot}$)& $3.02\times10^{12}$&  &$(7.48\pm{1.63})\times10^{11}$  \\ \relax
        IMF & $S2C^{1.8}$ &$S2C^{1.8}$ & \citet{Chabrier2003}   \\ 
        \end{tabular}\\ 
        
       
    	\begin{tabular}{|c|c|c|}\hline
    	\multicolumn{3}{|c|}{GRB 050915A}  \\ \hline \relax
    	Physical parameters & \citet{Perley2013a} & Best-fit(No AGN) \\ \hline \relax
        Stellar mass ($M_{\odot}$) & (3.67$^{+1.6}_{-1.04}$)$\times 10^{10}$  & (9.56$\pm{0.48}$)$\times 10^{10}$\\ \relax
        SFR ($M_{\odot}$ yr$^{-1}$) & 135.8$^{+63.1}_{-48.2}$ &$76\pm{4}$\\ \relax
        Total IR luminosity ($L_{\odot}$)& & $(8.58\pm{0.43})\times10^{11}$  \\ \relax
        IMF & $S2C^{1.8}$ &\citet{Chabrier2003}   \\ 
        \hline
        \end{tabular}\\ 
        S2C means that the original setting of IMF is \citet{Salpeter1955}. But they used some factor to transform to \citet{Chabrier2003} for a fair comparison.\\
        $^{1.8}$Divided by a factor of 1.8 from \citet{Salpeter1955} to \citet{Chabrier2003}\\
        $^M$Multiplied by a factor in \citet{Madau2014} (0.61 for stellar mass and 0.63 for SFR) from \citet{Salpeter1955} to \citet{Chabrier2003}.
        \end{flushleft}
    \end{table*}

    \section{Conclusion}
    \label{conclusion}
    
    With ALMA observations in the rest-frame FIR, we provide new estimates of SFRs for six GRB host galaxies. 
    The rest-frame FIR continuum were detected for the GRB080207 host
    with S/N=11, 
    the GRB060814 host with S/N$\sim$16, 
    the GRB070306 host with S/N$\sim$5, 
    the GRB081221 host with S/N$\sim12$, 
    the GRB071021 host with S/N$\sim$4,
    and the GRB050915A host with S/N$\sim$3.5.
    
    The most notable difference in this work compared with previous estimations without ALMA data is the SFR values.
    SFR with ALMA is $\sim2$ times larger than previous studies for the GRB080207 host.
    For host galaxies of GRB060814, GRB081221, GRB071021 and GRB050915A, the SFRs were overestimated in previous works, which are $\sim2$ to $\sim5$ larger than our estimations.
    The SED fitting analyses indicate SFRs of $397\pm20$, $56\pm9$, $38\pm2$, $56\pm3$, $90\pm5$ and $77\pm4$ $M_{\odot}$ yr$^{-1}$ for the GRB080207, GRB060814, GRB070306, GRB081221, GRB071021 and GRB050915A host galaxies, respectively. By adding ALMA data, this work derived a more reliable estimate of SFRs, which is important to understand GRB host galaxies 
    as a CSFR probe. 
    
    This change in SFR also leads to a more reliable estimate of the stellar population, and thereby, more reliable stellar masses of GRB host galaxies. For GRB050915A, $M_*$ is $\sim2.5$ larger than the values derived from previous paper. 
    The star-formation rate and other physical parameters are compared with previous literature in Tables \ref{tab12} and \ref{tab13}.
    
    With ALMA data, we improved our knowledge on the host galaxies as follows.
    We found one ULIRG within our sample: GRB080207 host galaxy ($L_{\rm IR}=(1.77\pm0.09)\times10^{12}L_{\odot}$). The IR luminosity of the other five host galaxies are between $10^{11}L{\odot}$ and $10^{12}L_{\odot}$, suggesting that they may be luminous infra-red galaxies (LIRGs).
    The best IMFs for our GRB host galaxies are \citet{Salpeter1955} for GRB080207 and GRB081221 host galaxies, 
    whereas \citet{Chabrier2003} for GRB060814, GRB070306, GRB071021 and GRB050915A host galaxies.

\section*{Acknowledgements}
We thank the anonymous referee for useful comments and constructive remarks on the manuscript. This paper makes use of the following ALMA data: ADS/JAO.ALMA\#2015.1.00927.S and 2016.1.00768.S. ALMA is a partnership of ESO (representing its member states), NSF (USA) and NINS (Japan), together with NRC (Canada), MOST and ASIAA (Taiwan), and KASI (Republic of Korea), in cooperation with the Republic of Chile. The Joint ALMA Observatory is operated by ESO, AUI/NRAO and NAOJ.
TG acknowledges the support by the Ministry of Science and Technology of Taiwan through grant 108-2628-M-007-004-MY3. TH and AYLO are supported by the Centre for Informatics and Computation in Astronomy (CICA) at National Tsing Hua University (NTHU) through a grant from the Ministry of Education of the Republic of China (Taiwan).
AYLO's visit to NTHU was hosted by Prof. Albert Kong and supported through the Ministry of Science and Technology of the ROC (Taiwan) grant 105-2119-M-007-028-MY3.
YYH is very grateful for the funding from The University Consortium of ALMA–Taiwan Summer Student Program.
This research has made use of the SVO Filter Profile Service (\url{http://svo2.cab.inta-csic.es/theory/fps/}) supported from the Spanish MINECO through grant AYA2017-84089.

\section*{Data availability}
\label{Data availability}
The data underlying this article will be shared on reasonable request to the corresponding author.

    \bibliographystyle{mnras}
    \bibliography{references}
    \bsp	
    \label{lastpage}
    \end{document}